\documentclass[pra,a4paper,showpacs,twocolumn,superscriptaddress,longbibliography]{revtex4-1}
\usepackage{amssymb}
\usepackage{amsmath}
\usepackage{amsfonts}
\usepackage{graphicx}
\usepackage{bm}
\usepackage{color}
\usepackage{multirow}
\usepackage{ulem}
\usepackage{relsize}

\DeclareMathOperator{\Tr}{Tr}

\definecolor{orange}{RGB}{255,127,0}

\begin{document}

\title{Helical edge transport in the presence of a magnetic impurity: the role of local anisotropy}

\author{Vladislav D.~Kurilovich}

\affiliation{Departments of Physics, Yale University, New Haven, CT 06520, USA}

\author{Pavel D.~Kurilovich}

\affiliation{Departments of Physics, Yale University, New Haven, CT 06520, USA}

\author{Igor S.~Burmistrov}

\affiliation{\hbox{L.~D.~Landau Institute for Theoretical Physics, acad. Semenova av. 1-a, 142432 Chernogolovka, Russia}}

\affiliation{Laboratory for Condensed Matter Physics, National Research University Higher School of Economics, 101000 Moscow, Russia
}

\affiliation{Institut f\"ur Theorie der kondensierten Materie, Karlsruhe Institute of Technology, 76128 Karlsruhe, Germany}

\affiliation{Institut f\"ur Nanotechnologie, Karlsruhe Institute of Technology, 76021 Karlsruhe, Germany}

\author{Moshe Goldstein}

\affiliation{Raymond and Beverly Sackler School of Physics and Astronomy, Tel Aviv University, Tel Aviv 6997801, Israel}

\date{\today} 

\begin{abstract}
Helical edge modes of 2D topological insulators are supposed to be protected from time-reversal invariant elastic backscattering. Yet substantial deviations from the perfect conductance are typically observed experimentally down to very low temperatures.
To resolve this conundrum we consider the effect of a single magnetic impurity with arbitrary spin on the helical edge transport. 
We consider the most general structure of the exchange interaction between the impurity and the edge electrons. We take into the account the local anisotropy for the impurity and show that it strongly  affects the backscattering current in a wide range of voltages and temperatures. We show that the sensitivity of the backscattering current to the presence of the local anisotropy is different for half-integer and integer values of the impurity spin. In the latter case the anisotropy can significantly increase the backscattering correction to the current. 
\end{abstract}

\maketitle

\section{Introduction}

Two-dimensional (2D) topological insulators are the subject of much recent interest due to their unique helical edge modes~\cite{Qi-Zhang,Hasan-Kane}. Strong spin-orbit coupling in these materials leads to spin-momentum locking of the edge electrons \cite{Kane-Mele, BHZ}, which has been detected experimentally in HgTe/CdTe quantum wells \cite{Konig2007,Roth2009,Gusev2011,Brune2012,Kononov2015}, and holds promise for numerous applications in spintronics.

In the presence of time-reversal symmetry, elastic backscattering of the helical electrons is forbidden. Hence, at low temperatures ballistic transport along the edge with quantized conductance of $G_0=e^2/h$ is expected. However, during the last decade this theoretical prediction was questioned by transport experiments 
in HgTe/CdTe~\cite{Konig2007,Nowack,Grabecki,Gusev2013,Gusev2014,Kvon2015} and 
InAs/GaSb~\cite{Knez2011,Suzuki2013,RRDu1,RRDu2,
Suzuki2015,Mueller2015,RRDu3,Mueller2017} quantum wells, as well as bismuth bilayers~\cite{Sabater2013} and WTe$_{2}$ monolayers~\cite{Cobden17,Jia2017,Herrero2018}.
Therefore, detailed studies of possible backscattering mechanisms at the helical edge are of the great importance. 
Many of the explanations raised in the literature involve significant electron-electron interactions at the edge~\cite{Xu2006,Schmidt2012,Lezmy2012,Maciejko2012,Yudson2013,Gornyi2014,Rod2015,Yudson2015,Glazman2016,Aseev2016,Meir2017,Hsu2017,Hsu2018,Yudson2017}.
However, since 2D topological insulator heterostructures typically contain nearby gates that effectively screen the interactions, these suggestions cannot satisfactorily account for all aspects of the experimental data. 

In the absence of the electron-electron interactions or time-reversal symmetry breaking, the ideal edge transport can still be affected at finite temperature by coupling to an impurity with its own quantum dynamics, e.g., a charge puddle that acts as an effective spin-${1}/{2}$ impurity~\cite{Goldstein2013, Goldstein2014}, or a quantum magnetic impurity with spin $S=1/2$,~\cite{Maciejko2009,Tanaka2011} or $S\geq 1/2$.~\cite{Cheianov2013,Kimme2016,Kurilovichi2017}2
\color{black}
The case $S>1/2$ offers a new prospect 
 with respect to spin 1/2, \color{black}
 since a local anisotropy term 
is generated due to the impurity's exchange interaction with nearby electrons~\cite{RKonig,Schiller}. This local 
anisotropy can dominate the dynamics of the impurity spin at low temperatures and voltages and, consequently, affect the 
helical edge transport. However, it has largely been overlooked till now.

In this work we theoretically study how the dc conductance of a noninteracting helical edge deviates from its ideal quantized value due to scattering off a single magnetic impurity with an arbitrary spin.  Contrary to previous works, we solve the problem for a generic structure of the matrix describing the exchange interaction between the edge electrons and the magnetic impurity. As a further generalization of the model, we take into account the presence of local anisotropy for the impurity spin. We discuss the cases of easy-plane anisotropy and easy-axis anisotropy, as well as of weakly non-uniaxial anisotropy. A physical case in point is a (001) CdTe/HgTe/CdTe quantum well contaminated by $\mathrm{Mn}^{2+}$ impurities, which possess spin $5/2$. Let us stress, however, that our theory is not restricted to this type of structure, and is suitable for the description of other 2D topological insulators as well.

We find that the backscattering current is sensitive to the parity of $2S$ and is strongly affected by the local anisotropy in a much wider range of voltage and temperature [see\ Eq.~\eqref{eq:condition}] than it is naively expected, especially for integer $S$. The current-voltage characteristics for the backscattering current possesses a rich phase diagram that is different for integer and for half-integer spin (see Figs.~\ref{Figure}a and \ref{Figure}c). Due to the presence of the local anisotropy, the 
dependence of the backscattering current on the voltage at low temperatures becomes strongly non-monotonous (see Figs.~\ref{Figure}b and \ref{Figure}d, as well as Secs. \ref{Sec:Overall1} and \ref{Sec:Overall2}). 


The outline of the paper is as follows. We start from formulation of the model in Sec.~\ref{Sec:Mod}. In Sec.~\ref{Sec:CC} we obtain a general expression for the backscattering current. The quantum master equation which describes the dynamics of the magnetic impurity coupled to the helical edge is derived in Sec.~\ref{Sec:QME}. The results for the backscattering current in the case of half-integer and integer spins are presented in Sec.~\ref{Sec:half-int} and \ref{Sec:IntS}, respectively.  We end the paper with conclusions (Sec.~\ref{Sec:Conc}). The details on some of the derivations are delegated to the Appendices. Throughout the text we use units in which $\hbar = k_B = -e=1$.
\color{black}

\section{Model\label{Sec:Mod}}

We start from the following Hamiltonian for a helical edge coupled to a magnetic impurity located at the position $y = y_{0}$ along the edge:
\begin{equation}
\label{eq: tot-H}
H=H_{\mathrm{e}}+H_{\mathrm{e-i}}+H_{\mathrm{i}}.
\end{equation}
Here $H_\mathrm{e}$ is the Hamiltonian of the edge electrons, $H_\mathrm{i}$ is the impurity Hamiltonian describing the local magnetic anisotropy, and $H_\mathrm{e-i}$ is the electron-impurity exchange interaction. We take $H_\mathrm{e}$ of the form
\begin{equation}
H_{\mathrm{e}}= i v\int dy\, \Psi^\dagger(y) \sigma_z \partial_y \Psi(y) ,
\end{equation}
where $v$ denotes the velocity of the edge states, $\Psi^\dagger$ ($\Psi$) is the creation (annihilation) operator of the edge electrons, and $\sigma_{x,y,z}$ are the Pauli matrices in the edge states spin space.

The exchange interaction between the helical electrons and the magnetic impurity is assumed to be local:
\begin{equation}
\label{eq: e-imp}
H_{\mathrm{e-i}}=\frac{1}{\nu}\mathcal{J}_{ij} S_i s_j(y_{0}),\quad s_j(y) = \frac{1}{2}\Psi^\dagger (y) \sigma_j \Psi (y) .
\end{equation}
Here $S_i$ denotes the components of the impurity spin operator, $\nu = 1/\left( 2\pi v \right)$ is the density of states per one edge mode, and the exchange couplings $\mathcal{J}_{ij}$ are real, dimensionless, and small $|\mathcal{J}_{ij}|\ll 1$.  It is worthwhile to mention that due to the presence of spin-orbit coupling in the 2D topological insulators the exchange matrix $\mathcal{J}_{ij}$ is not necessarily diagonal. For example, $\mathcal{J}_{ij}$ has four nonzero components, $\mathcal{J}_{xx}=\mathcal{J}_{yy}$,  $\mathcal{J}_{zz}$, and $\mathcal{J}_{xz}$, for an impurity in a HgTe/CdTe quantum well provided the interface inversion asymmetry is negligible~\cite{Otten2013,Kimme2016,Kurilovichi2017}. Taking the inversion asymmetry of HgTe/CdTe quantum wells~\cite{Dai2008,Konig2008,Winkler2012,Weithofer-Recher,Tarasenko2015,Durnev2016,Minkov2013,Minkov2016} into account, all components of the matrix $\mathcal{J}_{ij}$ become finite. Similar situation is expected to occur in other 2D topological insulators, e.g., InAs/GaSb quantum wells,  bismuth bilayers, and WTe$_2$ monolayers.
\color{black}


We note that the exchange interaction $\mathcal{J}_{ij}$ acquires Kondo-type renormalization~\cite{Zawadowski}. In what follows, we assume  
that the corresponding Kondo temperature is well below the relevant energy scales (related to the temperature, voltage, and local anisotropy), so that the renormalization of $\mathcal{J}_{ij}$ can be neglected (see Appendix \ref{app:sec:kondo}). This is typically justified physically: for example, for $\mathrm{Mn}^{2+}$ ion in a HgTe/CdTe quantum well $\mathcal{J}_{ij}\sim 10^{-3}$ \cite{Kurilovichi2017} and the corresponding Kondo temperature is extremely small (as compared to the energies accessible in transport experiments).\color{black}

Finally, the local anisotropy Hamiltonian is given by 
\begin{equation}
H_{\mathrm{i}}=\mathcal{D}_{qp}S_q S_p, 
\label{eq:Hi-st}
\end{equation}
where $\mathcal{D}_{qp}$ is a real symmetric matrix. To keep the discussion general, for the most part of the text we do not specify the mechanism behind the anisotropy and do not make restrictive assumptions on the relation between the coupling matrix $\mathcal{J}_{ij}$ and the anisotropy matrix $\mathcal{D}_{qp}$.  However, it should be noted that one of the possible sources of the anisotropy is the strong spin-orbit coupling in the topological insulator. \color{black}
Anisotropy of that type can be thought of as a result of the indirect exchange interaction of the magnetic impurity with itself, mediated by its coupling to both the bulk and the edge electronic states~\footnote{The higher order contributions to the local anisotropy 
associated with the indirect exchange interaction of the magnetic impurity with itself are smaller than the quadratic term $H_\mathrm{i} = \mathcal{D}_{qp} S_q S_p$ by additional powers of $\mathcal{J}_{ij}$, and hence are negligible. }. 
Assuming that all the elements $\mathcal{J}_{ij}$ are of the same order (we denote the corresponding value as ${J}$),
one may estimate $\mathcal{D}_{qp} \sim  J^2 |M| \Lambda$~\cite{Kurilovichi2016,Kurilovichi2017-0}, where $|M|$ is the bulk band gap (see Appendix \ref{App:Anis}). The dimensionless ultraviolet cut-off parameter $\Lambda$ is of order $ [v/\left(|M| a_{\mathrm{imp}}\right)]^3$.  Here $a_\mathrm{imp}$ is a typical range of the impurity potential. Using $a_\mathrm{imp} \sim 3$~nm, we find $\mathcal{D}_{qp} \sim 0.1$~K. 

With an appropriate $SO(3)$ rotation $R$ of the impurity spin, $\bm{S} = R \bm{S}^\prime $, it is always possible to simplify the local anisotropy (up to the constant energy shift) to the form,
\begin{equation}
\label{eq: s-anis}
H_{\mathrm{i}}=\mathcal{D}^\prime_{zz} {S_z^\prime}^2 + \mathcal{D}^\prime_{xx} {S_x^\prime}^2,
\end{equation}
with $\mathcal{D}\equiv|\mathcal{D}^\prime_{zz}|>|\mathcal{D}^\prime_{xx}|$.
The exchange matrix then becomes $\mathcal{J}^\prime = R^{-1}\mathcal{J}$. In what follows, we assume that the local anisotropy has the form \eqref{eq: s-anis} and thus omit the primes.

\section{Correction to the current\label{Sec:CC}} 

The helical nature of the edge states allows us to express the backscattering current, $\Delta I$, via the rate of change of the $z$-component of the total spin of the edge electrons: 
\begin{equation}
 \Delta I = \left\langle \frac{d}{d t}\int s_z(y)dy \right\rangle.
 \label{eq:dI:def}
 \end{equation}
Thus, if $S_z + \int s_z(y)dy$  is conserved, $ \Delta I=0$.~\cite{Tanaka2011,Goldstein2014}
This conservation can be broken by either
sufficiently anisotropic exchange $\mathcal{J}_{ij}$,~\cite{Kimme2016,Kurilovichi2017} or by the local anisotropy \eqref{eq: s-anis}, provided  $\mathcal{D}_{xx}$ is non-zero.

When a finite bias voltage $V$ is applied to the edge (we assume $V>0$), $s_z$ develops a non-zero expectation value $\nu V/2$. As a result, the Hamiltonian $H_\mathrm{e-i}$ acquires a non-zero mean-field shift: 
\begin{equation}
H_\mathrm{e-i}^{\mathrm{mf}} = \mathcal{J}_{iz} S_i V/2,
\label{eq:H:mf:e-i}
\end{equation}
 which acts as the effective Zeeman splitting for the magnetic impurity.
We denote eigenstates and energies of $H_\mathrm{i} + H_\mathrm{e-i}^{\mathrm{mf}}$ as $|\psi_a\rangle $ and $E_a$, respectively, where 
$a=S,S-1,\dots,-S$.

To the second order in $J$, we derived the following equation for the backscattering current (see Appendix \ref{app:QME}):
\begin{gather}
\Delta I =  \varepsilon_{zrj} \mathcal{J}_{ir}\mathcal{J}_{lk}\,\textrm{Im}\,  \sum_{cd} \mathcal{T}_V^{jk}(\omega_{cd}) \langle S_i \mathcal{S}_l^{cd} \rangle_S  .\label{eq:curr:corr-ex}
\end{gather}
Here $\varepsilon_{krj}$ is the Levi-Civita symbol, 
\begin{equation}
\mathcal{S}_l^{cd}=|\psi_c\rangle \langle \psi_c | S_l | \psi_d  \rangle \langle \psi_d|, 
\end{equation}
and  $\omega_{cd}=E_d-E_c$. The average $\langle \dots \rangle_S $ is taken over the reduced
density matrix of the magnetic impurity in the steady state, $\rho_S^{\mathrm{(st)}}$. The matrix
$\mathcal{T}_V(\omega)=\mathcal{T}_V^+(\omega) + \mathcal{T}_V^-(\omega)$
represents the spin-spin correlation  function of the edge electrons,
\begin{equation}
\mathcal{T}_V^\pm(\omega)=
\frac{\pi}{2}
\begin{pmatrix}
f(\omega\pm V) & \mp i f(\omega\pm V) & 0\\
\pm i f(\omega\pm V) & f(\omega\pm V) & 0\\
0 & 0 & f(\omega)
\end{pmatrix},
\label{eq:Tm:def}
\end{equation}
where we introduced $f(\omega)=\omega/[1 - \exp (-\omega/T)]$.

Below we will show that in many cases of interest it is possible to neglect $\omega_{cd}$ in the argument of the matrix $\mathcal{T}_V$ and use 
\begin{equation}
\mathcal{T}_V(0)
=\pi T
    \begin{pmatrix}
        \frac{V}{2T}\coth \frac{V}{2T}&-i\frac{V}{2T}&0\\
        i\frac{V}{2T}&\frac{V}{2T}\coth \frac{V}{2T}&0\\
        0&0&1
    \end{pmatrix}
\end{equation}
 instead.  If that is the case,  Eq. \eqref{eq:curr:corr-ex} may be substantially simplified \cite{Kurilovichi2017}:
\begin{align}    
\Delta I  = & \frac{\pi^2}{2} G_0 V \Bigl [\mathcal{X}_{j}\langle{S}_j\rangle_S \coth \frac{V}{2T} 
    \notag \\
    & - 2 \sum_{k=x,y}{\mathcal{J}}_{mk}{\mathcal{J}}_{nk}\langle S_m S_n\rangle_S\Bigr]  ,
    \label{sm:simp-curr}
\end{align}
where $\mathcal{X}_{j} = 2\varepsilon_{jkl}\mathcal{J}_{kx}\mathcal{J}_{ly}$.

\color{black}

\section{The quantum master equation\label{Sec:QME}}
 
In order to evaluate the backscattering current, it is necessary to determine the steady state density matrix $\rho_S^{\mathrm{(st)}}$. For this purpose we derived the Redfield equation \cite{BreuerPetruccione}, which governs the time evolution of the reduced density matrix $\rho_S$  (see Appendix \ref{app:QME}):
\begin{align}
    \frac{d\rho_S}{dt} & =-i\left[ {H}_\mathrm{i}+H_\mathrm{e-i}^{\mathrm{mf}}, \rho_S\right] \notag \\
    & + \frac{1}{2}\mathcal{J}_{rj}\mathcal{J}_{lk}  \Bigl (\sum_{cd}\mathcal{T}_V^{jk}(\omega_{cd})\left [\mathcal{S}_r^{cd}\rho_S, S_l\right ] 
    +\mathrm{h.c.}\Bigr )
  . \label{eq:denmat-ev}
\end{align}
The first term on the right hand side of Eq.~\eqref{eq:denmat-ev} describes the unitary dynamics of $\rho_S$, while the term quadratic in $\mathcal{J}$ accounts for Korringa-type relaxation due to weak coupling between the edge electrons and the impurity spin \cite{Korringa}. For $V=0$ the spin-spin correlation function 
$\mathcal{T}_{V=0}^{jk}(\omega) = \delta_{jk}\mathcal{T}_0(\omega)$ satisfies the detailed balance relation: $\mathcal{T}_0(-\omega)=e^{-\omega/T}\mathcal{T}_0(\omega)$. This leads to the thermal density matrix in the equilibrium steady state 
$\rho_S^\mathrm{(st)}\propto\sum_a \exp(-E_a/T)|\psi_a\rangle\langle \psi_a|$, and to the vanishing backscattering current.

\begin{figure*}[t]
\includegraphics[width=0.95\textwidth]{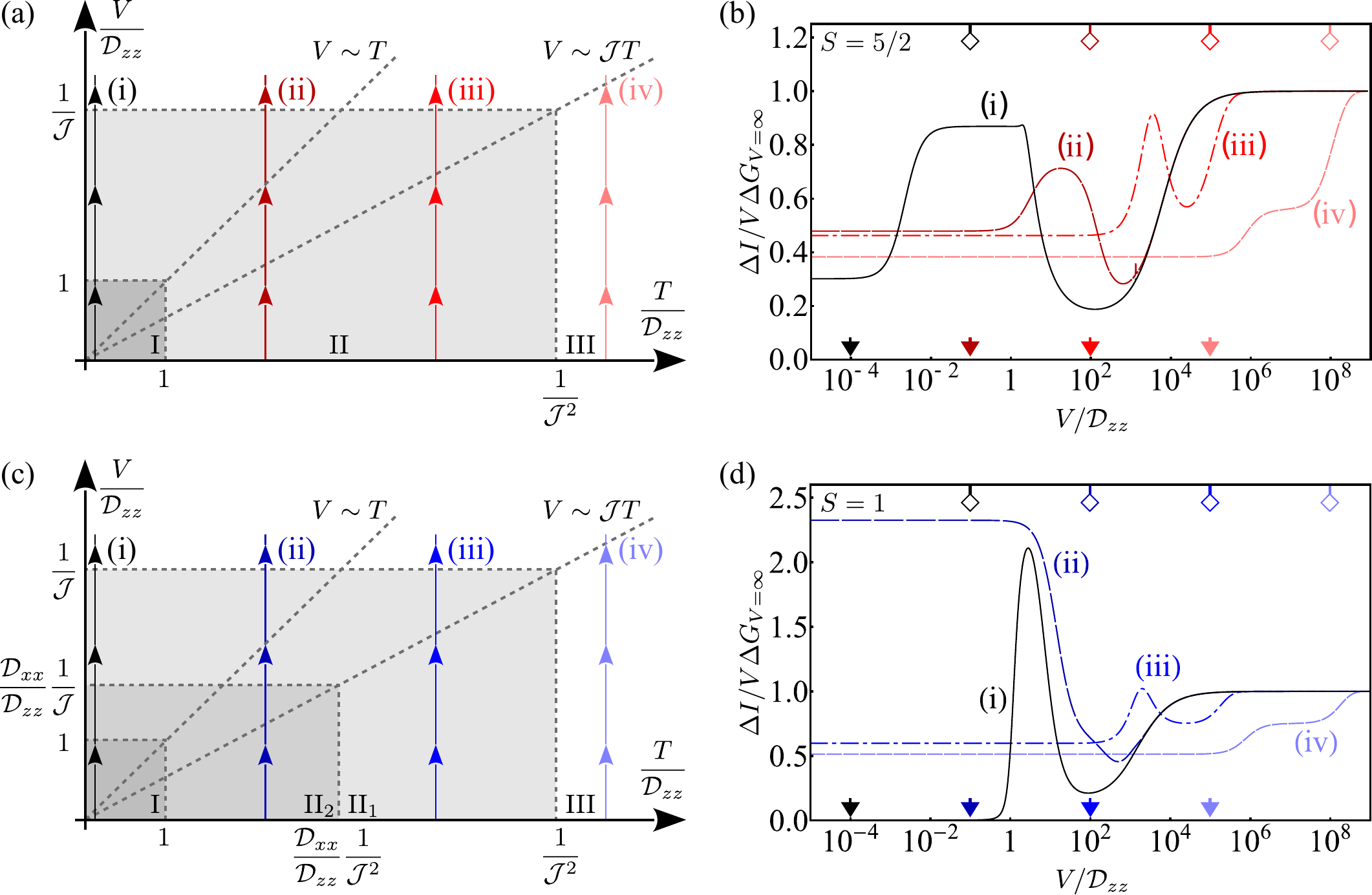}
\caption{Sketches of different regimes in $V$-$T$ plane for the half-integer (a) and integer (c) spin. The corresponding corrections to the current $\Delta I$ [obtained by numerical solution of Eqs. \eqref{eq:curr:corr-ex}-\eqref{eq:denmat-ev}] are plotted in panels (b) and (d) for $S=5/2$ and $S=1$, respectively; $\Delta I$ is normalized by $V\Delta G_{V = \infty}$, where $\Delta G_{V=\infty}$ denotes to the correction to the conductance in the large voltage limit.
The parameters are $\mathcal{D}_{zz}=1$, $\mathcal{D}_{xx}=0.01$, $\mathcal{J}_{xx}=\mathcal{J}_{yy}=10^{-3}$, $\mathcal{J}_{zz}=1.5 \mathcal{J}_{xx}$, $\mathcal{J}_{xz}=0.8 \mathcal{J}_{xx}$, $\mathcal{J}_{zx}=0.3 \mathcal{J}_{xx}$, and the temperature takes the values 
$T=10^{-1}$ (i), $T=10^2$ (ii), $T=10^5$ (iii), and $T=10^8$ (iv). Voltages of order $J T$ are denoted in panels (b) and (d) by down-pointing arrows (lower horizontal axes), whereas voltages equal to the temperature are marked by diamonds (upper horizontal axes). 
}
\label{Figure}
\end{figure*}

At non-zero voltage the unitary dynamics of $\rho_S$ is controlled by the effective Zeeman field
$\sim {J}V$ and the local anisotropy energy $\mathcal{D}$. 
The relaxation dynamics of $\rho_S$ is controlled by the Korringa rate $\tau_K^{-1}\sim{J}^2\max\{T,V,\mathcal{D}\}$. $\rho_{S}^{\mathrm{(st)}}$ depends on the relative magnitude of ${J}V$, $\mathcal{D}$, and $1/\tau_K$. This results in several distinct regimes in the $V$-$T$ plane for the steady state $\rho_S$ and for the backscattering current (see Fig.~\ref{Figure}).

In particular, in the regime $\max\{T,V\}\gg \mathcal{D}$ the relaxation is approximately insensitive to the local anisotropy [one can neglect the dependence of $\mathcal{T}_V^{jk}$ on $\omega_{cd}$ in Eq. 
\eqref{eq:denmat-ev}]. 

It is then possible to rewrite the master equation \eqref{eq:denmat-ev} {in the Gorini-Kossakowski-Sudarshan-Lindblad  form} (see Ref.~\onlinecite{Rivas} for a review)
\begin{align}
    \frac{d{\rho}_S}{dt}=& -i\left[{H}_\mathrm{i}+H_\mathrm{e-i}^{\mathrm{mf}},{\rho}_S\right] \notag \\
    & + {\eta}_{jk}\left({S}_{j}{\rho}_S{S}_{k}-\frac{1}{2}\{{\rho}_S, {S}_{k}{S}_{j}\}\right) ,
     \label{eq: ME simple}
\end{align}
    where \cite{Kurilovichi2017}
    \begin{equation} 
    {\eta}_{jk}= \left( \mathcal{J} \mathcal{T}_V(0) \mathcal{J}^T \right)_{jk} .
    \label{eq:eta:def}
\end{equation}

Nonetheless, even for $\max\{T,V\}\gg \mathcal{D}$ the local anisotropy cannot be always disregarded completely: due to the presence of the first term on the right hand side of Eq.~\eqref{eq:denmat-ev},
the anisotropy might still be crucial for the steady state density matrix, and thus, for the correction to the current. One can fully neglect $H_\mathrm{i}$ only if 
\color{black}
\begin{equation}
\max\{{J}^2T,{J}V\}\gg \mathcal{D} .
\label{eq:condition}
\end{equation}
Indeed, at large voltages, $V \gg \mathcal{D}/J$, the effective Zeeman field \eqref{eq:H:mf:e-i} dominates over the anisotropy, while at high temperatures, $T\gg \mathcal{D}/J^2$, the smearing $\sim \tau_K^{-1}$ of the energy levels of $H_\mathrm{i}$ due to the relaxation well exceeds the impurity level spacing.
In the absence of $H_\mathrm{i}$, Eq.~\eqref{eq:denmat-ev} has been analyzed recently by the present authors~\cite{Kurilovichi2017}. 
We note, however, that for spin-orbit coupling mediated anisotropy $\mathcal{D}/{J}^2$ is of order of the ultra-violet cut-off $\Lambda |M| \gg |M|$.
In that case, the anisotropy can be neglected only for temperatures  $T\gg \mathcal{D}/J^2 \gg |M|$  for which the current through the topological insulator is mainly carried by the bulk states.

In order to illustrate the importance of the local anisotropy for the backscattering current we consider both the case
of the \emph{easy-plane} anisotropy ($\mathcal{D}_{zz}>0$) 
and the case of the \emph{easy-axis} anisotropy ($\mathcal{D}_{zz}<0$). To simplify the discussion, we assume a clear hierarchy of scales $|\mathcal{D}_{zz}|\gg |\mathcal{D}_{xx}|$. 

The level structure of the total impurity Hamiltonian $H_\mathrm{i} + 
H_\mathrm{e-i}^{\mathrm{mf}}$, while being inherently important for determination of the backscattering current (as indicated by  Eqs.~\eqref{eq:curr:corr-ex} and \eqref{eq:denmat-ev}), differs qualitatively for integer and half-integer values of the impurity spin. {Therefore,} we consider these cases separately.

\section{Backscattering current for a half-integer spin $S$\label{Sec:half-int}}

In this section we consider in details the transport along a helical edge in the presence of a magnetic impurity with a half-integer spin $S$. We begin by inspecting the level structure of the Hamiltonian $H_{\rm i}+H_{\rm e-i}^{\rm mf}$.

\subsection{Level structure of the magnetic impurity}

We start from the case of no voltage applied to the edge of the topological insulator. If $\mathcal{D}_{xx}=0$ then the eigenstates of $H_\mathrm{i}$ are that of the $z$-projection of the impurity spin, $|\psi_{S_z}\rangle \equiv |S_z \rangle$, $S_z=+S,...,-S$. The energy levels are doubly degenerate: $E_{\pm S_z}= \mathcal{D}_{zz} S_z^2$. According to Kramers theorem, as long as half-integer spin is concerned, this degeneracy cannot be lifted by perturbations preserving time-reversal symmetry. Therefore, small $\mathcal{D}_{xx}$ leaves the degeneracy of energy levels intact while weakly altering the structure of the eigenstates. As a result, $\mathcal{D}_{xx}$ produces corrections to the backscattering current proportional to $\mathcal{D}_{xx}/\mathcal{D}_{zz}$ only, which we shall ignore below. 
Such approximation is well justified provided the matrix $\mathcal{J}$ has a generic form. For specific choices of $\mathcal{J}$, a small $\mathcal{D}_{xx}$ term in the Hamiltonian might still be important. For instance, if the electron-impurity exchange interaction preserves the total $z$-projection of angular momentum of the system, i.e., $\mathcal{J}_\mathrm{XXZ} = \textrm{diag}\,\{\mathcal{J}_\perp,\mathcal{J}_\perp,\mathcal{J}_z\}$, then $\mathcal{D}_{xx}$ is a sole source of backscattering. Hereinafter we concentrate on the generic case and neglect small $\mathcal{D}_{xx}$ for an impurity with a half-integer spin.

At finite voltage the mean-field part of the impurity Hamiltonian, $H_{\mathrm{e-i}}^{\mathrm{mf}} = \mathcal{J}_{iz} S_i V/2$, alters the level structure significantly. It effectively breaks time-reversal symmetry for the magnetic impurity, leading to voltage-dependent Zeeman-type splitting of the energy levels. The character of this splitting is different for the doublets $|\pm |S_z| \rangle$ with $|S_z|>1/2$ and $|S_z|=1/2$. For small $J V/\mathcal{D}_{zz}$, states with $|S_z|\neq 1/2$ are split trivially, i.e., the energy of the state $|S_z\rangle$ is shifted by $\mathcal{J}_{zz} S_z V/2$.
That is because the matrix element $\langle \pm S_z |S_{x/y}|S_z\rangle$ vanishes. For $S_z=\pm 1/2$ the matrix element $\langle-1/2|H_{\mathrm{e-i}}^{\mathrm{mf}}|1/2\rangle\neq 0$ and one has to solve the secular equation in order to extract the level shifts and the eigenstates. To the lowest order in $J V/\mathcal{D}_{zz}$ the corresponding effective Hamiltonian has the form:
\begin{equation}
    H_{\pm 1/2}^\mathrm{eff}= \frac{1}{4}\begin{pmatrix}
        \mathcal{D}_{zz} + V\mathcal{J}_{zz}&  \mathcal{J}_- V (S+\frac12) \\ \mathcal{J}_+ V (S+\frac12)    & \mathcal{D}_{zz}- V\mathcal{J}_{zz}
        \end{pmatrix},
\end{equation}
where $\mathcal{J}_{\pm}=\mathcal{J}_{xz}\pm i\mathcal{J}_{yz}$. Its eigenvalues are 
\begin{equation}
E_{\pm {1/2}}  = \frac{1}{4}\left(\mathcal{D}_{zz} \pm V\sqrt{\mathcal{J}_{zz}^2+(\mathcal{J}_{xz}^2+\mathcal{J}_{yz}^2)\Bigl(S+\frac12\Bigr)^2}\right) . 
\end{equation}
The respective eigenstates, $|\psi_{\pm 1/2} \rangle \equiv |\pm{1/2}^\prime\rangle$, are given by
\begin{gather}
    \begin{pmatrix}
    |+{1/2}^\prime\rangle\\|-{1/2}^\prime\rangle
    \end{pmatrix}=\begin{pmatrix}
    \cos\frac{\theta}{2}&e^{i\phi}\sin\frac{\theta}{2}\\
    -e^{-i\phi}\sin\frac{\theta}{2}&\cos\frac{\theta}{2}
    \end{pmatrix}\begin{pmatrix}
    |+{1/2}\rangle\\|-{1/2}\rangle
    \end{pmatrix},\notag \\
    \tan\theta = \left (S+\frac{1}{2}\right )\frac{\sqrt{\mathcal{J}^2_{xz}+\mathcal{J}^2_{yz}}}{\mathcal{J}_{zz}},\quad \tan \phi = \frac{\mathcal{J}_{yz}}{\mathcal{J}_{xz}}.
    \label{sm:rotation}
\end{gather}
This non-trivial modification of level structure  is very important in the high-energy regime. But before getting to it, we start the discussion of the transport properties of the helical edge with the low-energy regime of small temperatures and voltages.
 

\subsection{Low-energy transport}

\subsubsection{Easy-plane anisotropy}

First, we assume that the local anisotropy is of the \textit{easy-plane} type, $\mathcal{D}_{zz} > 0 $, and consider the regime $\mathrm{max}\{T,V\} \ll \mathcal{D}_{zz}$ (region I in Fig.~\ref{Figure}a). In that case, with exponential precision the impurity occupies the doubly-degenerate ground state subspace of $H_\mathrm{i}$ formed by the states with $S_z=\pm 1/2$, as can be inferred from the master equation \eqref{eq:denmat-ev}.
Therefore, it is possible to project the Hamiltonian \eqref{eq: tot-H} onto the doublet $|\pm 1/2\rangle$. The accuracy of such a projection is controlled by a small parameter $\max \{ T, V \}/\mathcal{D}_{zz} \ll 1$. Effectively, the projection maps the problem 
onto that of a  spin-$1/2$ impurity coupled to the edge states by the effective exchange matrix given by $\tilde{\mathcal{J}}_{xj}=(S+\frac{1}{2})\mathcal{J}_{xj}$, $\tilde{\mathcal{J}}_{yj}=(S+\frac{1}{2})\mathcal{J}_{yj}$, and $\tilde{\mathcal{J}}_{zj}=\mathcal{J}_{zj}$. The master equation \eqref{eq:denmat-ev} 
transforms into the following Gorini-Kossakowski-Sudarshan-Lindblad type equation for 
the $2\times 2$ reduced density matrix $\tilde{\rho}_S$ of the effective spin-$1/2$: 
\begin{equation}
    \label{eq: MEs12}
    \frac{d\tilde{\rho}_S}{dt}=-i\frac{V}{2}\tilde{\mathcal{J}}_{iz}\left[\tilde{S}_i,\tilde{\rho}_S\right]+ \tilde{\eta}_{jk}\left(\tilde{S}_{j}\tilde{\rho}_S\tilde{S}_{k}-\frac{1}{2}\{\tilde{\rho}_S, \tilde{S}_{k}\tilde{S}_{j}\}\right) .
    \end{equation}
Here the matrix $\tilde{\eta}_{jk}$ is given by Eq. \eqref{eq:eta:def} with $\tilde{\mathcal{J}}$ instead of $\mathcal{J}$ and $\tilde{S}_i = \sigma_i/2$ are effective spin-1/2 operators. Knowing the steady state density matrix  $\tilde{\rho}_S^{\left(\mathrm{st}\right)}$ one can calculate the correction to the current using Eq. \eqref{sm:simp-curr} with $\tilde{\mathcal{J}}$ and $\tilde{S}_i$ instead of $\mathcal{J}$ and $S_i$. 
A significant simplification comes from the relation $\sigma_i\sigma_j = \delta_{ij}+i\varepsilon_{ijk}\sigma_k$. Extracting the averages of spin operators in the stationary state from the master equation \eqref{eq: MEs12}, one finds 
\begin{gather}
   \Delta I = \frac{\pi^2}{4}     \left[\tilde{\mathcal{X}}^T \tilde{\Gamma}^{-1}\tilde{\mathcal{X}} \frac{V}{2T}\coth \frac{V}{2T} - \sum_{k=x,y}\tilde{\mathcal{J}}_{mk}\tilde{\mathcal{J}}_{mk}\right]G_0V,\notag \\
    \tilde{\Gamma}_{ij}=\frac{1}{\pi T}\left(\delta_{ij} \, \mathrm{tr}\, \tilde{\eta} - \frac{\tilde{\eta}_{ij}+\tilde{\eta}_{ji}}{2}+ V\varepsilon_{ijk} \tilde{\mathcal{J}}_{kz}\right) ,
    \label{eq:current:123}
\end{gather}
where $\tilde{\mathcal{X}}_j=2\varepsilon_{jkl}\tilde{\mathcal{J}}_{kx}\tilde{\mathcal{J}}_{ly}$. We stress that the backscattering current is of the second order in $J$ which is consistent with Fermi's golden rule.
Importantly, in the regime  $\mathrm{max}\{T,V\} \ll \mathcal{D}_{zz}$ the correction to the conductance $\Delta I / V$ saturates as function of voltage at $V\sim J T$ instead of the expected estimate $V\sim T$ \cite{Kurilovichi2017}.

\subsubsection{Easy-axis anisotropy}

Next, we consider the transport along the helical edge in the low-energy limit, $\max \left\{T, V\right\} \ll |\mathcal{D}_{zz}|$, assuming that the local anisotropy is of the \textit{easy-axis}  type, i.e., $\mathcal{D}_{zz}<0$.  In this regime, the impurity is constrained to occupy the subspace $|\pm S\rangle$. Consequently, to describe the backscattering current it is possible to project the Hamiltonian \eqref{eq: tot-H} onto the states $\left\{ |+S\rangle, |-S\rangle\right\}$. By doing so, we map the problem onto that of a spin-1/2 magnetic impurity which interacts with the edge electrons via a modified exchange matrix $\bar{\mathcal{J}}$. The components of $\bar{\mathcal{J}}$ are given by
$\bar{\mathcal{J}}_{xi}=\bar{\mathcal{J}}_{yi} = 0$, $\bar{\mathcal{J}}_{zi} = 2S \mathcal{J}_{zi}$ for $i=x,y,z$.
Then the master equation \eqref{eq:denmat-ev} can be reduced to the form of Eq. \eqref{eq: MEs12} with $\tilde{\mathcal{J}}$ substituted by $\bar{\mathcal{J}}$. Using the result \eqref{eq:current:123}, we find the following correction to the backscattering current:
\begin{equation}\label{sm:easy_axis}
\Delta I = -\frac{\pi^2}{4} (\bar{\mathcal{J}}_{zx}^2 + \bar{\mathcal{J}}_{zy}^2) G_0 V = -\pi^2 S^2 (\mathcal{J}_{zx}^2 + \mathcal{J}_{zy}^2) G_0 V .
\end{equation}
 

\subsection{Transport at high energies}

At $\max\{ T,V\}\gg |\mathcal{D}_{zz}| \gg\max\{ J^2 T,J V \}$  (region II in Fig. \ref{Figure}) the relaxation term in Eq.~\eqref{eq:denmat-ev} becomes independent of the anisotropy and the master equation simplifies to Eq.~\eqref{eq: ME simple}. Throughout this section we assume that  $V \gg J T$ which leads to the following relation between the energy scales in the problem: $|\mathcal{D}_{zz}|\gg J V \gg \tau_\mathrm{K}^{-1}$. This hierarchy allows us to exploit the rotating wave approximation \cite{Rivas} and to find the analytical expression for the steady state density matrix of the magnetic impurity. The latter is diagonal in the eigenstate basis $|\psi_m\rangle$ of $H_\mathrm{i}+H_{\rm e-i}^{\rm mf}$, 
\begin{equation}
\rho_S^{\rm (st)} = \sum_m p^\mathrm{(st)}_m |\psi_m\rangle \langle \psi_m | . 
\label{eq:stst}
\end{equation}
The coefficients $p^\mathrm{(st)}_m$ can be found by requiring the relaxation term in the master equation~\eqref{eq:denmat-ev} to be zero for such density matrix.
This condition can be written as
\begin{gather}\sum_n w_{m \leftarrow n} p_n = p_m \sum_n w_{n \leftarrow m},\notag \\
w_{n \leftarrow m} =  \eta_{ij} \langle \psi_n | S_i | \psi_m \rangle  \langle \psi_m | S_j | \psi_n \rangle .
\label{sm:clmark}
\end{gather}

As it was discussed previously, the states $|S_z\rangle$ with $|S_z|>1/2$ are approximate eigenstates of $H_{\mathrm{i}}^{\mathrm{full}}$. For the $|S_z|=1/2$ subspace the basis should be rotated as indicated by equation \eqref{sm:rotation}. Solving equation \eqref{sm:clmark}, we find that the stationary state the density matrix in the basis $\left\{|S\rangle,\dots,|{1/2}^\prime\rangle,|-{1/2}^\prime\rangle,\dots,|-S\rangle\right\}$ is given as follows:
\begin{equation}\label{sm:dm_he_hi}
\rho_S^{\left(\mathrm{st}\right)} \propto \mathrm{diag}\{{\vartheta}^S, \dots ,{\vartheta}^{3/2},a_1,a_2,b \vartheta^{-3/2},\dots,b \vartheta^{-S}\}.
\end{equation}
Here we introduce the real parameter 
\begin{equation}
\vartheta = 
\frac{\eta_{xx} +i\eta_{xy}-i(\eta_{yx}+i\eta_{yy})}{\eta_{xx} -i\eta_{xy}+i(\eta_{yx}-i\eta_{yy})}.
\label{eq:vartheta:def}
\end{equation}
The parameters $a_1$ and $a_2$ are given by
\begin{equation}
\begin{pmatrix}
a_1\\a_2
\end{pmatrix}=\frac{1}{\cos \theta}\begin{pmatrix}
\cos^2(\theta/2)-b \sin^2(\theta/2)\\b \cos^2 (\theta/2) - \sin^2(\theta/2)
\end{pmatrix} 
\end{equation}
with
$b = \mathrm{Tr}\left(\Theta\eta\right)/\mathrm{Tr}\left(\Theta^T \eta\right)$ .
Here $\Theta$ is a Hermitian $3\times3$ matrix whose elements are
\begin{gather}
	\Theta_{11}=\left (S+\frac12\right )^2\left(1-\frac{1}{2}\cos (2\phi)\sin^2 \theta\right)-\frac{1}{2}\sin^2 \theta,\notag \\
	 \Theta_{22}=\left (S+\frac12\right )^2\left(1+\frac{1}{2}\cos (2\phi)\sin^2 \theta\right)-\frac{1}{2}\sin^2 \theta,\notag	
	\\
    \Theta_{12}=i\left (S+\frac12\right)^2\left(\cos^2\theta+\frac{i}{2}\sin^2\theta\sin(2\phi)\right),\notag\\ 
    \Theta_{13}=-\left(S+\frac12\right)e^{i\phi}\sin\theta\cos\theta,\notag \\
     \Theta_{23}=i\left (S+\frac12\right)e^{i\phi}\sin\theta\cos\theta, \notag \\
    \Theta_{33}=\sin^2\theta .
\end{gather}

It is possible to find the correction to the current by substituting the obtained density matrix \eqref{sm:dm_he_hi} into Eq.~\eqref{sm:simp-curr}. {The asymptotic result \eqref{sm:dm_he_hi} is valid both for the \textit{easy-plane} and \textit{easy-axis} anisotropy.} 

At small voltage, 
$V \ll J T$, $\rho_S^\mathrm{(st)}$ has non-zero off-diagonal elements in the eigenbasis $|\psi_m\rangle$. This hinders analytic solution for the backscattering current.

\color{black}

In region III, $\max\{J^2 T, J V\}\gg |\mathcal{D}_{zz}|$,  (see Fig.~\ref{Figure}) the local anisotropy is completely irrelevant.  For $V\gg J T$ the solution for the steady state density matrix has a Gibbs form, with an effective temperature $T_{\rm eff}$ that depends on the ratio $V/T$.~\cite{Kurilovichi2017}

\subsection{The overall behavior of the backscattering current for a half-integer spin}
\label{Sec:Overall1}

The dependence of the backscattering conductance $\Delta G = \Delta I/V$ on voltage obtained from the numerical solution of Eq.~\eqref{eq:denmat-ev} for $S=5/2$ is shown in Fig.~\ref{Figure}(b) for several temperatures. Curve (i) corresponds to $T\ll \mathcal{D}_{zz}$.  The backscattering current at voltage $V\lesssim \mathcal{D}_{zz}$ (region I) at first rises and then, at $V\sim J T$, saturates to a plateau, 
in reminiscence of the spin-$1/2$ problem  \cite{Kurilovichi2017}.
At the boundary between regions I and II, $V \sim \mathcal{D}_{zz}$, the curve exhibits a cusp. It is associated with the emergence of transitions of the impurity to the excited states.
The wide minimum in curve (i) corresponds to 
region II in which $V\gg T$. At the crossover between regions II and III the minimum turns into the plateau corresponding to $\Delta I$ in the absence of the anisotropy. 
Curve (ii) {is plotted} for the temperature range $\mathcal{D}_{zz} \ll T \ll \mathcal{D}_{zz}/J$. At $V\sim JT$ the low-voltage plateau turns into the wide maximum and, then, into the minimum. 
The switching between the minimum and the maximum occurs at $V \sim T$.  At $V\sim D_{zz}/J$ the mean-field part of the impurity Hamiltonian $H_{\mathrm{e-i}}^{\mathrm{mf}}\sim J V$ becomes sufficiently large to significantly alter the structure of anisotropic energy levels. This leads to a transition between a minimum corresponding to region II and a high energy plateau corresponding to region III. A small peak in the backscattering conductance appears when two impurity levels come close together (a trace of this peak is also visible in curve (i)).
Curve (iii) {corresponds} to the temperature $\mathcal{D}_{zz}/J \ll T \ll  \mathcal{D}_{zz}/J^2$. It starts with a plateau at $V\ll JT$, which then turns into a maximum at $V \sim  \mathcal{D}_{zz}/J$,  associated with the crossover between regions II and III. In region III curve (iii) has a minimum corresponding to a {Gibbs-like} steady state with $\mathcal{D}_{zz}/J\ll V\ll T$.~\cite{Kurilovichi2017} At $V\gg T$ $\Delta G$ saturates at the plateau. Curve (iv) corresponds to $T\gg \mathcal{D}_{zz}/J^2$ so {that} the local anisotropy is irrelevant at any $V$. There are three plateaus in $\Delta G$ {positioned at} $V\ll JT$, $JT \ll V\ll T$, and $T\ll V$ respectively.


\section{Backscattering current for an integer spin $S$\label{Sec:IntS}}


This section is devoted to the transport along the helical edge in the presence of an impurity with integer spin $S$. Similarly to Sec.~\ref{Sec:half-int} we first discuss the level structure of $H_{\rm i}+H_{\rm e-i}^{\rm mf}$.

\subsection{Level structure of the magnetic impurity}

Contrary to the case of half-integer spin of the magnetic impurity, a small $\mathcal{D}_{xx}$ cannot be neglected for integer impurity spin. Let us start from the equilibrium limit, $V = 0$, and diagonalize $H_\mathrm{i}$ by treating the $\mathcal{D}_{xx}$ term in it as a perturbation. To do that, we notice that the energy levels of the unperturbed Hamiltonian $\mathcal{D}_{zz} S_z^2$ may be chosen to have a well-defined spin-$z$ projection $S_z$. Hence, for a given $S_z > 0$ a pair of levels $\left\{|+S_z\rangle,|-S_z\rangle\right\}$ is degenerate.  The presence of a finite $\mathcal{D}_{xx}$  lifts this degeneracy. The effective Hamiltonian which governs the splitting of $|\pm S_z\rangle$ doublet as well as its overall energy shift to the lowest non-vanishing order in $\mathcal{D}_{xx}$ is given by (the basis is $\left\{|+S_z\rangle,|-S_z\rangle\right\}$, where $S_z$ is assumed to be a positive integer) 
\begin{gather}
H^\mathrm{eff}_{\pm S_z} =
\begin{pmatrix}
\mathcal{D}_{zz}S_z^2+d_{S_z} & \Delta_{S_z}\\
\Delta_{S_z} & \mathcal{D}_{zz}S_z^2+d_{S_z}
\end{pmatrix} , \notag \\
\Delta_{S_z} = \mathcal{D}_{xx} \left( \frac{\mathcal{D}_{xx}}{\mathcal{D}_{zz}} \right)^{S_z - 1}
\frac{\prod\limits_{m=-S_z+2}^{S_z}\langle m | S_x^2 | m - 2 \rangle}{\prod\limits_{m=-S_z+2}^{S_z-2}(S_z^2-m^2)} ,
\end{gather}
where $d_{S_z} = \mathcal{D}_{xx} \left(S(S+1)-S_z^2\right)/2$.
As a result, the $|\pm S_z\rangle$ states split into a symmetric and antisymmetric combinations, $\left[|+S_z\rangle \pm |-S_z\rangle \right] /\sqrt{2}$, with energies $ \mathcal{D}_{zz}S_z^2 + d_{S_z} \pm \Delta_{S_z}$, respectively. We denote the corresponding energy gap as $\delta_{S_z} = 2|\Delta_{S_z}|$. As long as $S \sim 1$, the numerical factor in the expression for $\Delta_{S_z}$ is of order unity and therefore $\delta_{S_z} \sim |\mathcal{D}_{xx}| \left| \mathcal{D}_{xx} / \mathcal{D}_{zz} \right|^{S_z -1}$. In what follows we ignore the overall shift $d_{S_z}$ since it 
has no significant effect on the backscattering current in the regimes considered analytically.  

Finite voltage tends to split the doublets as well. In particular, if the anisotropy is purely uniaxial, $\mathcal{D}_{xx}=0$, the mean-field  electron-impurity interaction, $H^\mathrm{mf}_\mathrm{e-i}$, induces a splitting of $|\pm S_z\rangle$ into $|+S_z\rangle$ and $|-S_z\rangle$ with energies $E_{\pm S_z}=\mathcal{D}_{zz} S_z^2 \pm \mathcal{J}_{zz} S_z V/2$, respectively. When both finite $\mathcal{D}_{xx}$ and non-zero voltage are introduced, there is a competition between the two splitting mechanisms. If, for a given $S_z > 0$, $V \ll \delta_{S_z}/|\mathcal{J}_{zz}|$, then the $|\pm S_z\rangle$ doublet breaks into a symmetric and antisymmetric combinations with the energy separation $\simeq \delta_{S_z}$. In the opposite limit, $V \gg \delta_{S_z}/|\mathcal{J}_{zz}|$, the doublet splits trivially into the $|+ S_z\rangle$ and $|- S_z\rangle$ states, which are separated by an energy $\simeq \mathcal{J}_{zz} S_z V$. In what follows we assume that the matrix $\mathcal{J}$ is generic and, therefore, its element $\mathcal{J}_{zz}$ is of order of the typical value of $\mathcal{J}_{ij}$, i.e., $J$. Hence, the crossover between the two regimes happens at $V\sim \delta_{S_z}/J$.


\subsection{Low-energy transport}

\subsubsection{Easy-plane anisotropy}
To begin with, we assume the anisotropy of the \textit{easy-plane} type,  $\mathcal{D}_{zz}>0$, and consider the regime of the low-energy transport, $\max\{T,V\} \ll \mathcal{D}_{zz}$ (region I in Fig.~\ref{Figure}(c)). In this limit it is possible to neglect $\mathcal{D}_{xx}$ since it gives rise only to small corrections of order of $\mathcal{D}_{xx}/\mathcal{D}_{zz}\ll 1$ to the results for the backscattering conductance. We stress that such an approximation  is not valid at arbitrary energies as well as for the other sign of $\mathcal{D}_{zz}$.

For  $\max \{ T,V\} \ll \mathcal{D}_{zz}$ one can project the initial Hamiltonian~\eqref{eq: tot-H} onto the non-degenerate ground state of $H_\mathrm{i}$, which is the state with $S_z=0$ in the absence of $\mathcal{D}_{xx}$. This implies that the magnetic impurity becomes frozen and thus $\Delta I$ based on Eqs.~\eqref{eq:curr:corr-ex} and \eqref{eq:denmat-ev} is exponentially small in $T/\mathcal{D}_{zz}$. 
This exponentially-small correction is surpassed by the contribution from virtual transitions between the ground state and the pair of the lowest excited nearly degenerate states. These virtual transitions mediate the effective  interaction between the edge electrons with opposite helicity in the vicinity of the impurity. 
In order to estimate this effect, we project the electron-impurity interaction on the $|S_z = 0\rangle$ state to second order in $J$ and obtain the following low-energy Hamiltonian
\begin{align}
H_{\mathrm{e-e}}^{\mathrm{eff}} & = -\frac{1}{\mathcal{D}_{zz}} \mathcal{J}_{ik}\mathcal{J}_{jl} s_k s_l \sum_{S_z = \pm 1} \langle 0 | S_i | S_z \rangle \langle S_z | S_j | 0 \rangle 
\notag \\
& = -\frac{S(S+1)}{2\mathcal{D}_{zz}} \mathcal{J}_{ik}\mathcal{J}_{jl} s_k s_l(\delta_{ix}\delta_{jx}+\delta_{iy}\delta_{jy}) .
\label{sm:eff-int}
\end{align}
Here all operators $s_i$ are taken at the position of the magnetic impurity $y_0$. For the following, it is important to keep in mind that the electron-impurity interaction has a finite range $a_\mathrm{imp}$. Unless the finite range is taken into the consideration, the discussed correction to conductance due to virtual transitions vanishes, as dictated by the Pauli exclusion principle. To account for $a_\mathrm{imp} \neq 0$, we replace the electron spin density operators entering~\eqref{sm:eff-int} by
\begin{equation}
s_k \rightarrow \frac{1}{2}\int dy g(y-y_0) \psi^\dagger_\alpha (y) \sigma_k^{\alpha \beta} \psi_\beta (y),
\end{equation}
where $y_0$ is the position of the magnetic impurity at the edge and $g(y)$ is a symmetric smooth function satisfying $\int dy g(y) = 1$, $\int dy y^2 g(y) = a_\mathrm{imp}^2$. 

The effective electron-electron interaction \eqref{sm:eff-int}  mediates three types of two-particle scattering events,
\begin{align}\label{sm:proc}
(1) \quad &\left |s_{z,1} = \frac12,\:s_{z,2} = \frac12\right\rangle \rightleftarrows \left |s_{z,1} = - \frac12,\:s_{z,2} = - \frac12\right \rangle ,\notag\\
(2) \quad &\left |s_{z,1} = \frac12,\:s_{z,2} = \frac12\right \rangle \rightleftarrows \left |s_{z,1} = \frac12,\:s_{z,2} = - \frac12\right \rangle , \notag \\
(3) \quad &\left |s_{z,1} = - \frac12,\:s_{z,2} = - \frac12\right \rangle \rightleftarrows \left |s_{z,1} = \frac12,\:s_{z,2} = - \frac12\right \rangle,
\end{align}
where $s_z$ denotes the spin $z$-projection of helical electrons and $1,2$ indexes enumerate the interacting electrons. Process $(1)$ corresponds to the simultaneous backscattering of \textit{two} electrons. Processes $(2)$ and $(3)$ describe scattering events with one spin-flip. The Fermi golden rule may be employed in order to evaluate the associated rates~\cite{Schmidt2012,Lezmy2012}. A straightforward calculation yields the following estimates for the contributions to the backscattering current due to the processes of type $(1)$, $(2)$, and $(3)$ in \eqref{sm:proc}:
\begin{align}
\Delta I_{1} \sim & - \frac{S^2(S+1)^2}{\mathcal{D}_{zz}^2 v^4} G_0 V \sum_{k,j,p,r,m,n=x,y}\mathcal{J}_{jk}\mathcal{J}_{pr}\mathcal{J}_{jm}\mathcal{J}_{pn} \notag \\
\times & \left(\delta_{kr}\delta_{mn} - \varepsilon_{kr} \varepsilon_{mn} \right) \bigl (\max\left\{T, V\right\}\bigr )^6 a_\mathrm{imp}^4  \label{sm:virp1}
\end{align}
and
\begin{align}
\Delta I_{2\:\&\:3} \sim & - \frac{S^2(S+1)^2}{\mathcal{D}_{zz}^2 v^4} G_0 V \sum_{k,j,p=x,y} \mathcal{J}_{jk}\mathcal{J}_{pk}\mathcal{J}_{jz}\mathcal{J}_{pz} \notag \\
\times & \bigl (\max\left\{|\mu|,T,V\right\}\bigr )^2 \bigl (\max\left\{T, V\right\}\bigr )^4 a_\mathrm{imp}^4 ,\label{sm:virp2}
\end{align}
where $\varepsilon_{jk}=\varepsilon_{jkz}$, $|\mu| = vk_F$ is the chemical potential. In the limit $\max \left\{T,V \right\}\ll |\mu|$, processes with one electron spin-flip give a parametrically dominant contribution to the backscattering current at small energies, $\left|\Delta I_{1}\right| \ll \left|\Delta I_{2\:\&\:3}\right|$.

\subsubsection{Easy-axis anisotropy} 

Next, we consider the transport along the helical edge in the low-energy limit, assuming that the local anisotropy is of the \textit{easy-axis}  type, i.e., $\mathcal{D}_{zz}<0$. Similarly to the case of the impurity with half-integer spin, in this regime the dynamics of the magnetic impurity is restricted to the subspace  $\left\{ |+S\rangle, |-S\rangle\right\}$. The projection of \eqref{eq: tot-H} on this subspace maps the problem onto that of a spin-1/2 coupled to helical electrons by the exchange matrix $\bar{\mathcal{J}}$ with the components
\begin{equation}
\bar{\mathcal{J}}_{xi}=\bar{\mathcal{J}}_{yi} = 0,\quad \bar{\mathcal{J}}_{zi} = 2S \mathcal{J}_{zi},\quad i=x,y,z.
\end{equation}

Provided that $\delta_S \ll \max \left\{ T, V\right\} \ll |\mathcal{D}_{zz}|$, we recover Eq.~\eqref{sm:easy_axis} for the backscattering current in full analogy with the case of half-integer spin of the impurity.

If $\max \left\{ T, V\right\} \ll \delta_S$, the impurity is frozen in its ground state, i.e., either $\left[ |+S\rangle - |-S\rangle\right]/\sqrt{2}$ or $\left[ |+S\rangle + |-S\rangle\right]/\sqrt{2}$ depending on 
the sign of $\mathcal{D}_{xx}$. Therefore, the leading contribution to the backscattering current is produced by virtual transitions of the impurity to the lowest excited state. The evaluation of the corresponding correction to the helical edge conductance with the help of the Fermi golden rule yields $\Delta I =  \Delta \bar{I}_{1} + \Delta \bar{I}_{2\:\&\:3}$, where
\begin{gather}
\Delta \bar{I}_{1}  \sim  - \frac{S^4 a_\mathrm{imp}^4 }{\delta_S^2 v^4}  \Bigl(\sum_{k=x,y}\mathcal{J}^2_{zk}\Bigr)^2\bigl (\max\left\{T, V\right\}\bigr )^6  G_0 V,\label{sm:virp3}\\
\Delta \bar{I}_{2\:\&\:3}  \sim  - \frac{S^4 a_\mathrm{imp}^4 }{\delta_S^2 v^4}  \sum_{k=x,y} \mathcal{J}^2_{zk}\mathcal{J}^2_{zz} \bigl( \max{\left\{|\mu|,T,V\right\}}\bigr )^2 \notag \\
  \times \bigl (\max\left\{T, V\right\}\bigr )^4 G_0 V.\label{sm:virp4}
\end{gather}
{We note that the character of the backscattering current for the \textit{easy-axis} anisotropy in the regime $\max \left\{ T, V\right\} \ll \delta_S$ is qualitatively similar to that for the \textit{easy-plane} anisotropy in the low-energy limit, $\max \left\{ T, V\right\} \ll \mathcal{D}_{zz}$. Indeed, the dependence of $\Delta I$ on voltage and temperature is similar between Eqs.~\eqref{sm:virp3}, \eqref{sm:virp4} and Eqs.~\eqref{sm:virp1}, \eqref{sm:virp2}. Yet, the expressions \eqref{sm:virp3} and  \eqref{sm:virp4} are parametrically different from \eqref{sm:virp1} and  \eqref{sm:virp2} and are determined by different combinations of the dimensionless coupling constants $\mathcal{J}_{ij}$.}

\subsection{Transport at high energies}

\color{black}

Contrary to the case of half-integer spin, the behavior of $\Delta I$ in the region II, $\max\{T,V\}\gg \mathcal{D}_{zz}\gg\max\{J^2 T,J V \}$,  is sensitive to the presence of non-zero $\mathcal{D}_{xx}$. The competition between the effective Zeeman splitting $H_\mathrm{e-i}^{\mathrm{mf}}$, the Korringa rate $1/\tau_K$, and the splittings $\delta_{S_z}$ leads to crossovers at $\max\{J^2 T, J V\} \sim \delta_{S_z}$ with $S_z=1,...,S$ (see Fig.~ \ref{Figure}(c)). In the subsequent sections we explore the character of backscattering in the region II, separately considering the limits of strongly smeared impurity levels, $V \ll JT$ (i.e, the Korringa rate $\tau_K^{-1}$ is much larger than the Zeeman-type splitting $\sim JV$), and the the limit of well separated impurity levels, $V \gg JT$.
 
We note that the results presented below are applicable for both the \emph{easy-plane} anisotropy and the \emph{easy-axis} anisotropy.

\subsubsection{Strongly smeared energy levels, $V \ll J T$}

In the regime $V \ll J T$ the steady state density matrix of the magnetic impurity is close to equipartitioning, 
\begin{equation}
\left(\rho_S^\mathrm{(eq)}\right)_{S_z,S_z^\prime} =\langle S_z|\rho_S^{\mathrm{(eq)}}|S_z^\prime\rangle= \frac{1}{2S+1}\delta_{S_z,S_z^\prime}.
\end{equation} 
The deviations of $\rho^\mathrm{(st)}_S$ from $\rho_S^\mathrm{(eq)}$ are proportional to $V/T$. Therefore, we expand
\begin{equation}\label{sm:dec}
    \rho_S^\mathrm{(st)}=\rho_S^\mathrm{(eq)} + \frac{V}{T} \delta \rho_S  + ...,\quad\eta_{ij} = \eta^{(0)}_{ij} -i \frac{V}{T} \eta^{(1)}_{ij}  + ...,
\end{equation} and examine the structure of $\delta \rho_S$. It is worthwhile to mention that $\eta^{(0)}_{ij}$ is a symmetric matrix, whereas $\eta_{ij}^{(1)}$ is antisymmetric. Substituting the decompositions \eqref{sm:dec} into Eq.~\eqref{eq: ME simple} and projecting the resulting equation onto the states $|S_z\rangle$ and $|S_z^\prime\rangle$ we find
\begin{gather}
(S_z^2 - S_z^{\prime 2})\left(\delta \rho_S\right)_{S_z,S_z^\prime} = -\frac{\mathcal{D}_{xx}}{\mathcal{D}_{zz}} \left( \left[S_x^2, \delta \rho_S \right]\right)_{S_z,S_z^\prime}
\notag \\
-i\frac{\eta^{(0)}_{ij}}{\mathcal{D}_{zz}}\left(S_i \delta \rho_S S_j - \frac{1}{2}\left\{S_j S_i, \delta \rho_S \right\}\right)_{S_z,S_z^\prime} \notag \\
+\varepsilon_{ijk}\frac{\eta^{(1)}_{ij}}{\mathcal{D}_{zz}} \left(S_k\right)_{S_z,S_z^\prime}.
\label{sm:ME-dec}
\end{gather}
Notice that we disregarded the mean-field part of the electron-impurity interaction Hamiltonian $H_\mathrm{e-i}^\mathrm{mf}$ in Eq.~\eqref{sm:ME-dec}. It is justified since we consider the regime $V \ll J T$. In Eq.\eqref{sm:ME-dec}, $|\eta^{(0,1)}_{ij}/\mathcal{D}_{zz}| \sim J^2 T /|\mathcal{D}_{zz}| \ll 1$ and $|\mathcal{D}_{xx}/\mathcal{D}_{zz}|\ll 1$ are small parameters. Hence, it is possible to neglect the components $(\delta \rho_S)_{S_z,S_z^\prime}$ with $|S_z| \neq |S_z^\prime|$ as compared to those with $|S_z|=|S_z^\prime|$ and, consequently, solve \eqref{sm:ME-dec} in the diagonal subspace $|S_z| = |S_z^\prime|$. An immediate consequence of such separation is that $|\langle S_{x,y}\rangle_S | \ll |\langle S_z \rangle_S|$. This observation, as well as the fact that the steady state density matrix is close to $\rho_S^\mathrm{(eq)}$, allows us to reduce the expression for the backscattering current \eqref{sm:simp-curr} to 
\begin{gather}
    \Delta I = \pi^2\frac{S(S+1)}{3} \left[\frac{T}{V}  \frac{3{\mathcal{X}}_z }{ S(S+1)} \langle{S}_z\rangle_S - g\right] G_0 V,\notag \\
     g = \sum_{k=x,y}{\mathcal{J}}_{mk}{\mathcal{J}}_{mk}. \label{sm:curr-dxx}
\end{gather}

Next we note that in the regime $V\ll J T$ a hierarchy of \textit{temperatures} arises: the backscattering current is sensitive to whether the Korringa relaxation rate, $\tau_K^{-1} \sim J^2 T$, surpasses $\delta_{S_z}$ with different $S_z > 0$.


If $\delta_1 \ll J^2 T \ll |\mathcal{D}_{zz}|$, (region II$_1$ in Fig. \ref{Figure}(c)), the doublets with all possible $|S_z|$ are well smeared. Thus, it is possible to disregard the $\mathcal{D}_{xx}$ term in the right-hand side of Eq.~\eqref{sm:ME-dec}. From the remaining system of equations for the diagonal components of $\delta \rho_S$ we find
\begin{gather}
    \left(\delta \rho_S\right)_{S_z,S_z} = S_z \frac{2\varepsilon_{jkz} \eta_{jk}^{(1)}}{(\eta_{xx}^{(0)}+\eta_{yy}^{(0)})(2S+1)} \notag \\
    = S_z \frac{\mathcal{X}_z}{(\Gamma_0)_{zz}(2S+1)},
\end{gather}
where $\Gamma_0 = \mathrm{Tr} \left(\mathcal{J}\mathcal{J}^T\right)-\mathcal{J}\mathcal{J}^T$.
Therefore,
\begin{equation}
\langle S_z\rangle_S = \frac{S(S+1)}{3}\frac{V}{T} \frac{\mathcal{X}_z}{(\Gamma_0)_{zz}},
\end{equation}
and
\begin{equation}\label{sm:inter_curr_low}
    \Delta I = - \pi^2\frac{ S(S+1) }{3} \left(g - \frac{\mathcal{X}_z^2}{ (\Gamma_0)_{zz}}\right) G_0V.
\end{equation}
 
In the regime $\delta_2 \ll J^2 T \ll \delta_1$, (region II$_2$ in Fig. \ref{Figure}(c)), the doublet $|\pm S_z\rangle$ with $S_z = 1$ is well split, whereas all other doublets are smeared. Solving equation \eqref{sm:ME-dec} to the leading order in $J^2 T / \delta_1$ while keeping in mind that $J^2 T \gg \delta_2$, we obtain the following expression for the diagonal components of $\delta \rho_S$:
\begin{equation}\label{sm:dm_000}
\left(\delta \rho_S \right)_{S_z, S_z} = 
\begin{cases} \frac{(S_z - 1) \mathcal{X}_z}{\left(\Gamma_0\right)_{zz}(2S+1)},\quad &S_z > 1,\\
0,\quad &|S_z|\leq 1,\\
\frac{(S_z + 1) \mathcal{X}_z}{\left(\Gamma_0\right)_{zz}(2S+1)},\quad &S_z < -1.
\end{cases}
\end{equation}
Then we find
\begin{equation}
\langle S_z \rangle = \frac{2}{3} \frac{(S^2-1)S}{2S+1}\frac{\mathcal{X}_z}{\left(\Gamma_0\right)_{zz}}\frac{V}{T},
\end{equation}
and, consequently,
\begin{equation}\label{sm:inter_curr_low_2}
\Delta I = - \pi^2 \frac{ S(S+1) }{3}\left(g - \frac{2(S-1)}{2S+1}\frac{\mathcal{X}_z^2}{\left(\Gamma_0\right)_{zz}}\right) G_0V.
\end{equation}

The expressions for the backscattering current may be derived in other regions $\delta_{S_z + 1} \ll J^2 T \ll \delta_{S_z}$, $S_z>1$, in a similar manner.


\subsubsection{Well separated impurity levels, $V \gg J T$}

Provided that $V \gg J T$, the splitting of each doublet $|\pm S_z\rangle$ with $S_z > 0$ exceeds the level smearing due to relaxation. In this regime the rotating wave approximation may be used to describe the dynamics of the impurity. The steady state density matrix acquires the diagonal form \eqref{eq:stst} with coefficients satisfying Eq. \eqref{sm:clmark}.

The next steps are sensitive to the precise structure of the impurity levels $|\psi_m\rangle$. As discussed above, this structure strongly depends on the ratio between the mean-field interaction $\sim JV$ and the splittings $\delta_{S_z}$.

In particular, if $\delta_1 \sim |\mathcal{D}_{xx}| \ll JV \ll |\mathcal{D}_{zz}|$, (region II$_1$ in Fig. \ref{Figure}(c)), the splittings of all doublets are determined predominantly by $H_\mathrm{e-i}^{\mathrm{mf}}$,  and each $|\pm S_z \rangle$ pair simply splits into $|+S_z\rangle$ and $|-S_z\rangle$.  Enumerating the energy levels as $|\psi_m\rangle = |S_z = m\rangle$, $m=S,...,-S$, we reduce Eq.~\eqref{sm:clmark} to a \textit{tridiagonal} system of differential equations, $\mathcal{M}_{m,n} p_n^{\rm (st)} = 0$, with a  matrix $\mathcal{M}_{m,n}$ which has the following non-zero elements:
\begin{align}
\mathcal{M}_{m\pm 1,m} & = (\eta_{xx} \pm i \eta_{xy} ) \mp i (\eta_{yx} \pm i \eta_{yy}) \notag \\
& \times (S(S+1)-m(m\pm 1))/4,\\
\mathcal{M}_{m,m} & = - \sum_{s=\pm 1} (\eta_{xx} + i s \eta_{xy} ) - i s (\eta_{yx} + i s \eta_{yy}) 
\notag \\
& \times (S(S+1)-m(m+s ))/4.
\end{align}
Then the steady state solution  can be readily found explicitly:
\begin{equation}\label{sm:vartheta}
p^{\mathrm{(st)}}_m =\mathcal{N} \vartheta^{m},
\end{equation}
where $\vartheta$ is defined in Eq. \eqref{eq:vartheta:def} and $\mathcal{N}$ is a normalization constant which ensures that $\sum_m p^{\mathrm{(st)}}_m =1$. Alternatively, this result may be rewritten in the operator form:
\begin{equation}\label{sm:gibbsSz}
\rho^\mathrm{(st)}_S = \mathcal{N} \vartheta^{S_z}.
\end{equation}
Therefore, the steady state density matrix has the Gibbs form in the eigenbasis of $S_z$. The explicit expression \eqref{sm:gibbsSz} for the density matrix allows for a straightforward evaluation of the backscattering current with the help of Eq. \eqref{sm:simp-curr}.

In the case $\delta_2 \sim |\mathcal{D}_{xx}| \left| \mathcal{D}_{xx} / \mathcal{D}_{zz} \right| \ll J V \ll |\mathcal{D}_{xx}| \sim \delta_1$, (region II$_2$ in Fig. \ref{Figure}(c)), the level structure is somewhat more complicated. The doublets $|\pm S_z\rangle$ with $S_z > 1$ are split by the mean-field interaction into $|+S_z\rangle$ and $|-S_z\rangle$ states, whereas the doublet with $S_z = 1$ is split by $\mathcal{D}_{xx}$ into symmetric and antisymmetric superpositions $\left[ |+1\rangle \pm |-1\rangle \right]/\sqrt{2}$. The modification of the eigenstates structure alters the matrix $\mathcal{M}$, and it loses its tridiagonal form. Nonetheless, the analytic solution for the steady state density matrix can still be found. It has a \emph{non-Gibbs} form, and is given by
\begin{equation}
\rho_S^\mathrm{(st)}\propto\textrm{diag}\bigl\{\vartheta^{S-1},\dots,\vartheta,1,1,1,\vartheta^{-1},\dots,\vartheta^{-S+1}\bigr\} .
\label{eq:dms:anis}
\end{equation}
We emphasize that the result~\eqref{eq:dms:anis} implies equal probabilities of states with $S_z=1,0,-1$, i.e., the impurity spin in the presence of non-zero $\mathcal{D}_{xx}$ and voltage tends to behave partially as \emph{a classical spin}.

We note that for $S=1$ the steady state density matrix is $\rho_S^\mathrm{(st)}\approx \textrm{diag}\{1/3,1/3,1/3\}$ for $\max(V,T)\gg |\mathcal{D}_{zz}|$ and $\max(JV,J^2T)\ll \delta_1$ (see Eq.~\eqref{sm:dm_000} and Eq.~\eqref{eq:dms:anis}). This implies that in a broad range of $V$ and $T$ the backscattering current  is given by
\begin{equation}
\Delta I_{S=1} = -(2\pi^2/3) g G_0 V .
\label{eq:spin1cond}
\end{equation}
In accordance with general expectations,  $\Delta I_{S=1}$ remains finite even for the exchange interaction matrix close to  $\mathcal{J}_\mathrm{XXZ} = \textrm{diag}\,\{\mathcal{J}_\perp,\mathcal{J}_\perp,\mathcal{J}_z\}$ due to the presence of finite $\mathcal{D}_{xx}$ \footnote{We note that in {the} case $\mathcal{J}=\mathcal{J}_\mathrm{XXZ}$ the anisotropy mediated by spin-orbit coupling (see discussion after \eqref{eq:Hi-st}) is uniaxial, $\mathcal{D}_{xx} = 0$.}. Interestingly, for $\mathcal{J}=\mathcal{J}_\mathrm{XXZ}$ and for $S=1$ the backscattering current does not contain smallness in $\mathcal{D}_{xx}/\mathcal{D}_{zz}$ in contrast to the case of half-integer spin of the impurity. Thus, Eq.~\eqref{eq:spin1cond} implies a parametrically large enhancement of the backscattering current due to the presence of non-zero $\mathcal{D}_{xx}$ for $\max(V,T)\gg |\mathcal{D}_{zz}|$ and $\max(JV,J^2T)\ll \delta_1$. The discussed enhancement is not specific for $S=1$, it is present for all integer $S>1$.

In principle, every interval of voltages $\delta_{S_z + 1} \ll JV \ll \delta_{S_z}$, $S_z > 0$, may be analyzed in a similar fashion. 

\color{black}

\subsection{The overall behavior of the backscattering current for an integer spin}
\label{Sec:Overall2}
The backscattering conductance as a function of voltage obtained from the numerical solution of Eq.~\eqref{eq:denmat-ev} for $S=1$ and for different $T$ is shown in Fig.~\ref{Figure}(d). Curve (i) {corresponds to} $T\ll \mathcal{D}_{zz}$.  The backscattering current in region I,
$V\lesssim \mathcal{D}_{zz}$, is exponentially small. The
evolution of $\Delta I$ near the maximum corresponds to the crossover from region I to region II$_2$ and then to region II$_1$. The wide minimum in curve (i) {is associated with the structure of} the steady state solution $\rho_S^{\rm (st)}$ in the region II$_1$. Switching from the minimum to the plateau around $V\sim \mathcal{D}_{zz}/J$ corresponds to the crossover between regions II$_1$ and III. Curve (ii) {is plotted} for the temperature $\mathcal{D}_{zz} \ll T \ll \delta_1 /J^2$. 
Around $V\sim \mathcal{D}_{xx}/\mathcal{D}_{zz}\mathcal{J}$, $\Delta I$ drops down from the low-voltage plateau due to the crossover between the regions II$_2$ and II$_1$. The minimum in curve (ii) corresponds to region II$_1$ in which $V\gg T$.
The crossover between regions II and III at $V\sim  D_{zz}/J$ causes switching from the minimum to the high-voltage plateau.
We emphasize that the low-voltage plateaus of the curves (ii) in Fig.~\ref{Figure}(b) and Fig.~\ref{Figure}(d) are different due to the effect of $\mathcal{D}_{xx}$ in the case of integer spin.
Curves (iii) and (iv) in Fig.~\ref{Figure}(d) are plotted for temperatures obeying $\delta_1/J^2 \ll T \ll  \mathcal{D}_{zz}/J^2$ 
and $\mathcal{D}_{zz}/J^2\ll T$, respectively. Since at these temperatures the effect of $\mathcal{D}_{xx}$ on $\Delta I$ is negligible, these curves are qualitatively very similar to the corresponding curves in Fig.~\ref{Figure}(b).

\section{Conclusions\label{Sec:Conc}}

To summarize, we presented the results of a detailed study of the dc transport along the helical edge in the presence of a magnetic impurity. We considered a realistic model with an arbitrary value of the impurity spin $S$, with a general form of the exchange matrix, and with a local anisotropy. We found that the backscattering current is strongly affected by the local anisotropy at voltage and temperature satisfying $\max\{ J^2 T, J V\} \ll \mathcal{D}$,  
for which the energy splittings of the impurity states due to the local anisotropy Hamiltonian $H_\mathrm{i}$ are non-negligible. We revealed that the local anisotropy makes the backscattering current sensitive to the parity of $2S$. For integer $S$ we found that the local anisotropy can significantly increase the correction to the current in a certain range of $V$ and $T$. 

Our results predict that the backscattering correction to the \textit{linear} conductance is almost independent of the temperature down to very low temperatures (well below $\mathcal{D}$) for all cases except 
the case of integer spin and the easy-plane anisotropy for which  strong temperature dependence ($\sim T^4$) sets at temperature of the order of $\mathcal{D}$. The backscattering correction which is independent of $T$ in wide temperature range is consistent with experimental findings.  In the case of HgTe/CdTe quantum wells, for temperatures $T\gg \mathcal{D}$,  a typical backscattering correction to the linear conductance due to a single impurity can be estimated as~\cite{Kurilovichi2017} $|\Delta G(0)|/G_0 \sim 10^{-4} \div 10^{-3}$.

Let us assume that there is a finite 1D density $n_{\rm imp}$ of magnetic impurities at the helical edge of length $L$. Then, neglecting correlations in the backscattering processes on different magnetic impurities, the total edge resistance $R$ is simply the sum of the individual single-impurity resistances, $\delta R = |\Delta G|/G_0^2$. Then we find that the total resistance is proportional to the length of the edge, $R = L n_{\rm imp} |\Delta G|/G_0^2$, in accordance with experimental observations of Refs.~\cite{Knez2011,Grabecki,Gusev2014,Mueller2015,Mueller2017}. 
This estimate for $R$ holds under assumptions that the impurities are uncorrelated, $1/n_\mathrm{imp} > L_T = v/T$. Now, in a HgTe/CdTe quantum well at $T = 4.2$~K the thermal length $L_T$ is of the order of a micron (taking $v=0.4$~eV$\cdot$nm). On the other hand, resistive behavior with resistance of the order of $h/e^2$ typically starts for samples which are a few micrometers long. Since each impurity contributes $\Delta G/G_0 \lesssim 10^{-3}$, as mentioned above, the 1D distance between impurities along the edge should be $1/n_\mathrm{imp} \lesssim 10^{-2}$~$\mu$m, well below $L_T$. Thus, one would need to go beyond the independent-impurity approximation, which has not yet been done for fully-anisotropic $S \geq 1/2$ impurities (see Refs.\cite{Maciejko2012,Yudson2013,Cheianov2013,Yudson2015}). We leave that for future work.

We also note that our results can be extended to take into account the effect of electron-electron interactions within the Luttinger liquid description of the helical edge. In this case one can use the quantum master equation \eqref{eq:denmat-ev} and the expression for the current \eqref{eq:curr:corr-ex} but with the kernel $\mathcal{T}_V^{jk}(\omega)$ modified by the electron-electron interaction in a way described in Refs.~\cite{Probst,Lezmy2012,Glazman2016}. 
 
Finally, our theoretical results indicate that  the backscattering current can serve as a probe for the level structure of the magnetic impurities contaminating the helical edge. Our theory can thus provide a basis  for a systematic experimental study of rare magnetic impurities through the transport along the helical edge.

\begin{acknowledgements}
We thank Y. Gefen  for fruitful collaboration on the initial stage of this project, and him, L.I. Glazman, and B. Rosenow for very useful discussions. Hospitality by Tel Aviv University, the Weizmann Institute of Science, the Landau Institute for Theoretical Physics, and the Karlsruhe Institute of Technology is gratefully acknowledged. The work was partially supported by the Russian Foundation for Basic Research under Grant No.~17-02-00541, the program ``Contemporary problems of low-temperature physics'' of Russian Academy of Science,  the Alexander von Humboldt Foundation, the Israel Ministry of Science and Technology (Contract No.~3-12419), the Israel Science Foundation (Grant No.~227/15), the German Israeli Foundation (Grant No.~I-1259-303.10), the US-Israel Binational Science Foundation (Grant No.~2016224), and a travel grant by the BASIS Foundation.
\end{acknowledgements}

\appendix

\section{Kondo renormalization}\label{app:sec:kondo}

In this Appendix we discuss the renormalization of the electron-impurity coupling constants $\mathcal{J}_{ij}$. As long as the running energy scale $E$ is larger than the local anisotropy scale, $E\gg \mathcal{D}$, the one-loop renormalization group (RG) equations for $\mathcal{J}_{ij}$ have the following form \cite{Kurilovichi2017,Zawadowski}
\begin{equation}
\frac{d\mathcal{J}_{ij}}{d \tau} = \frac{1}{2}\varepsilon_{ikp} \varepsilon_{jmn} \mathcal{J}_{km}\mathcal{J}_{pn},\quad \tau = \ln \left(|M|/E\right).
\end{equation}
To simplify the system of equations, we perform a singular value decomposition of the coupling matrix: $\mathcal{J}=R_{<}{\lambda}R_{>}$. Here the matrices $R_{>}$ and $R_{<}$ are orthogonal, $R_>$ and $R_<\in SO(3)$, and ${\lambda}=\mathrm{diag}\left(\lambda_1,\lambda_2,\lambda_3\right)$. For the RG flow of the singular values we find
\begin{equation}
\frac{d\lambda_1}{d\tau}=\lambda_2\lambda_3,\quad \frac{d\lambda_2}{d\tau}=\lambda_3\lambda_1,\quad \frac{d\lambda_3}{d\tau}=\lambda_1\lambda_2,
\end{equation}
while the matrices $R_{>}$ and $R_{<}$ do not flow. When two of $\lambda_i$s are zero, the remaining coupling stays constant with the change of the energy scale. If two couplings are equal, e.g. $\lambda_1=\lambda_2\neq 0$, $\lambda_3\leq 0$, and $|\lambda_1|\leq|\lambda_3|$, then $\lambda_1$ goes to zero while $\lambda_3$ saturates at some finite value as $E$ is decreased. In all other cases, a finite Kondo energy scale $T_K$ exists at which $\lambda_i$s blow up. {The Kondo energy may be estimated as $T_K \sim |M| \exp \left( - 1/\mathcal{J}^0 \right)$, where $\mathcal{J}^0$ is a dimensionless parameter of order of $J$ at $\tau = 0$.}  As $T_K$ is approached, the coupling constants tend to the manifold $|\lambda_1|=|\lambda_2|=|\lambda_3|$ with $\lambda_1\lambda_2\lambda_3>0$.

Physically, the running energy scale is always determined by either the temperature or the voltage. Hence, the above analysis is applicable provided $\max \left\{T,V\right\}\gg \mathcal{D}$, whereas at lower energies the RG equations alter significantly \cite{Zitko}. Throughout the main text of the article we assume that $T_K$ is much smaller than $\max \left\{T,V, \mathcal{D}\right\}$ and therefore the renormalization of the exchange couplings can be neglected at the relevant energy scales. 
{This assumption is typically well-justified. For instance, for a Mn$^{2+}$ impurity in a topological insulator based on CdTe/HgTe/CdTe quantum well with width of $7 \:\mathrm{nm}$ the typical value of the exchange coupling $J(\tau = 0)$ is of order of $10^{-3}$ \cite{Kurilovichi2017}. Thus  $T_K$ is extremely small.}

\section{The local magnetic anisotropy \label{App:Anis}}

In this Appendix, we demonstrate how the local anisotropy of the magnetic impurity can be generated by the exchange interaction between  the impurity  and the electron states (both bulk and edge ones) in a 2D topological insulator. To simplify derivation we consider a CdTe/HgTe/CdTe quantum well and neglect the inversion asymmetry. 
In order to describe the electronic states in this structure we employ the linearized Bernevig-Hughes-Zhang Hamiltonian,
\begin{equation}\label{sm:H}
\mathcal{H}_\mathrm{e} = \begin{pmatrix}
M & v k_+ & 0 & 0\\
v k_- & -M & 0 & 0\\
0 & 0 & M & - v k_- \\
0 & 0 & - v k_+ & -M
\end{pmatrix},
\end{equation}
where $M$ is a band gap, $v$ will turn out to be the edge states velocity, and $k_{\pm} = k_x \pm i k_y$. The Hamiltonain $\mathcal{H}_{\mathrm{e}}$ is written in the basis of spatially quantized states $\left\{ \left| E_1,+\right\rangle, \left| H_1,+\right\rangle, \left| E_1,-\right\rangle, \left| H_1,-\right\rangle \right\}$ (see Ref.~\onlinecite{BHZ} for details). Notice that the Hamiltonian $\mathcal{H}_{\mathrm{e}}$  is rotationally invariant: for simplicity, we disregarded symmetry-lowering interface inequivalence \cite{Tarasenko2015} in the discussion of the magnetic anisotropy. To account for the presence of the edge in the system we follow the approach of Ref.~\onlinecite{VolkovPankratov} and assume that the gap is a function of $x$-coordinate such that the band inversion is realized at $x=0$, i.e., for $x < 0 $ $M(x)$ is a negative constant, whereas $ M ( x > 0 ) \to + \infty $.

The Hamiltonian of the local electron-impurity exchange interaction is given by $\mathcal{H}_{\mathrm{e-i}} = \mathfrak{J}^q S_q \delta (\bm{r} - \bm{r}_0)$, where $\mathfrak{J}^{x,y,z}$ are $4\times 4$ matrices in the basis $\left\{ \left| E_1,+\right\rangle, \left| H_1,+\right\rangle, \left| E_1,-\right\rangle, \left| H_1,-\right\rangle \right\}$, summation over $q$ is assumed, $\bm{r}_0$~is a position of the impurity in the quantum well, and $\bm{S}$ is the impurity spin operator.  An analysis based on the $k\cdot p$ method yields \cite{Kimme2016,Kurilovichi2016}:
\begin{equation}\label{sm:imp}
\mathfrak{J}^qS_q = 
\begin{pmatrix}
J_1 S_z & -iJ_0 S_+ & J_{m} S_{-} &0\\
iJ_0S_{-} & J_2 S_z &0 &0\\
J_{m} S_{+} &0 & -J_1 S_z & - iJ_0 S_{-}\\
0 & 0 &  iJ_0 S_+ & -J_2 S_z   
\end{pmatrix},
\end{equation}
where $S_\pm = S_x \pm i S_y$ and $J_0$, $J_1$, $J_2$, and $J_m$ are real parameters that depend on the microscopic details of the exchange interaction as well as on the structure of the envelop functions of the spatially quantized states $\left|E_1,\pm \right\rangle$ and $\left|H_1,\pm \right\rangle$. For the sake of universality, throughout this section we assume that all $\mathfrak{J}^q$s have a generic form and do not refer to the explicit form \eqref{sm:imp}.

The local magnetic anisotropy is generated by the indirect exchange interaction of the magnetic impurity with itself. A zero temperature expression for the indirect exchange, evaluated to second order in the coupling parameters $\mathfrak{J}^q$, is given by Eq. \eqref{eq:Hi-st} with 
\begin{equation}\label{sm:ani}
\mathcal{D}_{qp} =  \frac{1}{2}\int \frac{d\epsilon}{2\pi} \Tr \mathcal{G} (i\epsilon , \bm{r}_0, \bm{r}_0) \mathfrak{J}^q  \mathcal{G} (i\epsilon , \bm{r}_0, \bm{r}_0) \mathfrak{J}^p  .
\end{equation}
The Matsubara Green's function $\mathcal{G} (i\epsilon , \bm{r}_1, \bm{r}_2)$, which enters this expression, can be conveniently expressed as a sum over states,
\begin{equation}\label{sm:lehm}
\mathcal{G} (i\epsilon , \bm{r}_1, \bm{r}_2) = \sum_j \frac{\psi_j (\bm{r}_1) \psi_j^\dagger (\bm{r}_2)} {i\epsilon - E_j + \mu},
\end{equation}
where $\psi_j(\bm{r})$ are the eigenstates of $\mathcal{H}_{\mathrm{e}}$, $E_j$ denotes the corresponding energies, and $\mu$ is the chemical potential. The representation \eqref{sm:lehm} allows to divide the Green's function into two parts, $\mathcal{G} = \mathcal{G}_\mathrm{bulk} + \mathcal{G}_\mathrm{edge}$, where $\mathcal{G}_\mathrm{bulk}$ incorporates the sum over the bulk states  and $\mathcal{G}_\mathrm{edge}$ includes the sum over the edge states. As a result, it is possible to split the anisotropy matrix $\mathcal{D}_{qp}$ into three terms of different nature:
\begin{equation}\label{sm:iei}
\mathcal{D}_{qp} = \mathcal{D}_{qp}^{\mathrm{bulk}} + \mathcal{D}_{qp}^{\mathrm{edge}} + \mathcal{D}_{qp}^{\mathrm{int}},
\end{equation}
where
\begin{gather}
\mathcal{D}_{qp}^{\mathrm{bulk\:(edge)}} = \frac{1}{2}\int \frac{d\epsilon}{2\pi} \Tr \mathcal{G}^{\mathrm{bulk\:(edge)}} (i\epsilon , \bm{r}_0, \bm{r}_0) \mathfrak{J}^q \notag \\
\times  \mathcal{G}^{\mathrm{bulk\:(edge)}} (i\epsilon , \bm{r}_0, \bm{r}_0) \mathfrak{J}^p ,
 \\ 
\mathcal{D}_{qp}^{\mathrm{int}} = \frac{1}{2}\int \frac{d\epsilon}{2\pi} \Tr \mathcal{G}^{\mathrm{bulk}} (i\epsilon , \bm{r}_0, \bm{r}_0) \mathfrak{J}^q \mathcal{G}^{\mathrm{edge}} (i\epsilon , \bm{r}_0, \bm{r}_0) \mathfrak{J}^p
\notag \\  + (q\leftrightarrow p).
\end{gather}
The explicit structure of the eigenstates is required to estimate each of the contributions in Eq.~\eqref{sm:iei}. In the described setting the edge states wave functions are given by
\begin{align}
\psi_{\mathrm{edge}}^{\uparrow}(k_y,\bm{r}) & =
\begin{pmatrix}
1\\ i\\ 0\\ 0
\end{pmatrix} \theta (-x) \frac{e^{-|x|/\xi}}{\sqrt{2\pi\xi}} e^{ik_y y}, 
\notag \\
 \psi_{\mathrm{edge}}^{\downarrow}(k_y,\bm{r}) & =
\begin{pmatrix}
0\\ 0\\ 1\\ -i
\end{pmatrix} \theta (-x) \frac{e^{-|x|/\xi}}{\sqrt{2\pi\xi}} e^{ik_y y},
\end{align}
where $\theta(x)$ is a Heaviside step function and $\xi = |M|/v$. They are characterized by a dispersion which is exactly linear in the model \eqref{sm:H}, $E_\mathrm{edge}^{\uparrow/\downarrow}(k_y)= \mp v k_y$.

Due to the presence of the edge, the bulk states acquire a more complicated structure as compared to that in the infinite sample (for the details, see Ref.~\onlinecite{Kurilovichi2017-0}):
\begin{align}
\psi_{\mathrm{bulk}, \uparrow}^{\pm}(\bm{r}) & =
\begin{pmatrix}
\pm f^\pm_x(\pm\bm{k})\\
\pm i f^\mp_x(\pm\bm{k}) \\
0\\
0
\end{pmatrix}\frac{e^{ik_y y}}{2\pi},\notag\\
\psi_{\mathrm{bulk}, \downarrow}^\pm(\bm{r})&=
\begin{pmatrix}
0\\
0\\
\mp f^\pm_x(\mp\bm{k})\\
\pm if^\mp_x(\mp\bm{k})
\end{pmatrix}\frac{e^{ik_y y}}{2\pi}.
\end{align}
The dimensionless functions $f^{\pm}_x (\bm{k})$ which enter the expressions above are 
\begin{equation}
f^{\pm}_x(\bm{k}) = \theta (-x) \frac{\left( v k_\pm \pm i\left( {\cal{E}}(k)\mp|M|\right)\right) e^{ik_x x} + \mathrm{c.c.}}{2\sqrt{{\cal{E}}(k)({\cal{E}}(k)+vk_y)}}.
\end{equation}
Here $\mathcal{E}(k) = \sqrt{M^2 + v^2 k^2}$. The corresponding energies are $E_{\mathrm{bulk},\uparrow}^{\pm}(\bm{k}) = E_{\mathrm{bulk},\downarrow}^{\pm}(\bm{k}) = \pm \mathcal{E}(k)$. 

When the magnetic impurity is far away from the edge, $|x| \gg \xi$, $\mathcal{D}_{qp}^{\mathrm{edge}}$ and $\mathcal{D}_{qp}^{\mathrm{int}}$ are exponentially suppressed in comparison with the bulk contribution, $\mathcal{D}_{qp}^{\mathrm{bulk}}$, while the latter equals $\mathcal{D}_{qp}^{\mathrm{bulk}} = -\Lambda^{\mathrm{bulk}}_{\infty} |M|^3 \Tr\left( \mathfrak{J}^q \mathfrak{J}^p \right)/v^4$ with the dimensionless factor
\begin{gather}
 \Lambda^{\mathrm{bulk}}_{\infty} \sim \int \frac{d\epsilon}{2\pi} \frac{d^2\bm{k}_1}{(2\pi)^2}\frac{d^2\bm{k}_2}{(2\pi)^2} \frac{v^4 \epsilon^2/|M|^3}{\left(\epsilon^2 + \mathcal{E}(k_1)^2\right)\left(\epsilon^2 + \mathcal{E}(k_2)^2\right)} 
 \notag \\
 \sim \frac{v^4}{|M|^3}\int \frac{d\bm{k}_1}{(2\pi)^2} \frac{d\bm{k}_2}{(2\pi)^2} \frac{1}{\mathcal{E}(k_1) + \mathcal{E}(k_2)}.
\end{gather}
The integral diverges at high momenta and should be regularized. The ultraviolet cut-off momentum $k_{\mathrm{uv}}$ is determined by the  size of the impurity potential $a_{\mathrm{imp}}$, $k_{\mathrm{uv}} \sim 1/a_\mathrm{imp}$. Then one estimates $\Lambda^{\mathrm{bulk}}_{\infty} \sim \left( \xi / a_\mathrm{imp} \right)^3$.

When the impurity is exactly at the edge, $x=0$, $\mathcal{D}_{qp}^{\mathrm{bulk}}$, $\mathcal{D}_{qp}^{\mathrm{edge}}$, and $\mathcal{D}_{qp}^{\mathrm{int}}$ have a similar matrix structure, although they feature parametrically different numeric prefactors:
\begin{align}
\mathcal{D}_{qp}^{\mathrm{bulk}} & = -\Lambda^{\mathrm{bulk}}_0 \frac{|M|^3}{v^4}\Tr \left( \mathcal{P} \mathfrak{J}^q \mathcal{P} \mathfrak{J}^p  \right),\notag \\
 \mathcal{D}_{qp}^{\mathrm{edge}} & = -\Lambda^{\mathrm{edge}}_0 \frac{|M|^3}{v^4}\Tr \left( \mathcal{P} \mathfrak{J}^q \mathcal{P} \mathfrak{J}^p  \right),\notag \\
  \mathcal{D}_{qp}^{\mathrm{int}} & = -\Lambda^{\mathrm{int}}_0 \frac{|M|^3}{v^4}\Tr \left( \mathcal{P} \mathfrak{J}^q \mathcal{P} \mathfrak{J}^p  \right), 
\end{align}
where the matrix $\mathcal{P}$ equals 
\begin{equation}
\mathcal{P} =
\begin{pmatrix}
1 & -i & 0 & 0\\
i & 1 & 0 & 0\\
0 & 0 & 1 & i\\
0 & 0 & -i & 1
\end{pmatrix},
\end{equation}
and the prefactors are given by
\begin{align}
\Lambda^{\mathrm{bulk}}_0 & = \int \frac{d\epsilon}{2\pi} \frac{d^2\bm{k}_1}{(2\pi)^2}\frac{d^2\bm{k}_2}{(2\pi)^2} \frac{4v^4\epsilon^2/|M|^3}{\left(\epsilon^2 + \mathcal{E}(k_1)^2\right)\left(\epsilon^2 + \mathcal{E}(k_2)^2\right)}
\notag \\
& \times \frac{v^4 k_{x,1}^2 k_{x,2}^2}{\mathcal{E}(k_1)\mathcal{E}(k_2)(\mathcal{E}(k_1)+v k_{y,1})(\mathcal{E}(k_2)+v k_{y,2})}  \notag\\
&\sim \frac{v^4}{|M|^3}\int \frac{d^2\bm{k}_1}{(2\pi)^2} \frac{d^2\bm{k}_2}{(2\pi)^2} \frac{1}{\mathcal{E}(k_1) + \mathcal{E}(k_2)}\sim \left( \xi / a_\mathrm{imp} \right)^3, \notag \\
\Lambda^{\mathrm{edge}}_0 & = \int \frac{d\epsilon}{2\pi} \frac{dk_{y,1}}{2\pi}\frac{d k_{y,2}}{2\pi} \frac{v^2 \epsilon^2/|M|}{\left(\epsilon^2 + v^2k_{y,1}^2\right)\left(\epsilon^2 + v^2k_{y,2}^2\right)}\notag \\
& \sim \frac{1}{|M|}\int d\epsilon \sim 1, \notag \\
\Lambda^{\mathrm{int}}_0 & = \int \frac{d\epsilon}{2\pi} \frac{d^2\bm{k}_1}{(2\pi)^2}\frac{d k_{y,2}}{2\pi} \frac{4v^3 \epsilon^2/M^2}{\left(\epsilon^2 + \mathcal{E}(k_1)^2\right)\left(\epsilon^2 + v^2 k_{y,2}\right)}\notag \\
& \times \frac{v^2 k_{x,1}^2}{\mathcal{E}(k_1)(\mathcal{E}(k_1)+v k_{y,1})} \sim \frac{1}{|M|}\int d\epsilon \sim 1.
\end{align}
In the last two estimates we have taken into account that the energy of the edge states is limited by $\epsilon_\mathrm{uv} \sim |M|$.

The size of the impurity potential $a_\mathrm{imp}$ can be reliably estimated to be of order of several lattice spacings, $\sim 1$~nm. For example, for a manganese ion $\mathrm{Mn}^{2+}$ embedded into CdTe lattice we find $a_\mathrm{imp} \simeq a_B \varepsilon_\mathrm{CdTe}m_\mathrm{e}/2 m_\mathrm{CdTe}\simeq 3$~nm, where $a_B$ is the Bohr radius, $m_\mathrm{CdTe} \simeq 0.1m_\mathrm{e}$ is the electron band effective mass in CdTe, $m_\mathrm{e}$ is the bare electron mass, and $\varepsilon_\mathrm{CdTe}\simeq 10$ is the dielectric constant of CdTe.  At the same time, $\xi\simeq 40$~nm for the realistic parameters of a CdTe/HgTe/CdTe quantum well with width of 7~nm (see \cite{BHZ} for details). Hence, $\xi/a_\mathrm{imp}\gg 1$ can be considered a large parameter. It means that the anisotropy is mainly induced by the interaction between the impurity and the bulk states, $\mathcal{D}_{qp}^{\mathrm{bulk}}\gg \mathcal{D}_{qp}^{\mathrm{edge}},\:\mathcal{D}_{qp}^{\mathrm{int}}$. This conclusion is independent of the distance $|x|$ between the impurity and the edge. It is worthwhile to mention that $\Lambda_0^{\mathrm{bulk}}$ is of the same order as $\Lambda_{\infty}^{\mathrm{bulk}}$. Therefore, as the impurity is displaced from the edge into the bulk, the local anisotropy roughly preserves its value, while its matrix structure gradually changes from $\Tr \left(\mathcal{P} \mathfrak{J}^q \mathcal{P} \mathfrak{J}^p  \right)$ to $\Tr \left(\mathfrak{J}^q \mathfrak{J}^p \right)$ on a length scale $\Delta x \sim \xi$. 

Finally, we note that for the impurity located precisely at the edge, $\mathcal{D}_{qp}^\mathrm{bulk\:(edge,\:int)}$ can be equivalently rewritten as
\begin{equation}
\mathcal{D}_{qp}^\mathrm{bulk\:(edge,\:int)} = - 2\pi^2\Lambda_0^\mathrm{bulk\:(edge,\:int)} |M| \left(\mathcal{J}\mathcal{J}^T\right)_{qp},
\end{equation}
where $\mathcal{J}$ is the matrix of dimensionless couplings introduced in the main text. 

\section{Derivation of the quantum master equation and the expression for the current\label{app:QME}}

In this Apendix, we derive the quantum master equation which governs the behavior of the reduced density matrix of the magnetic impurity and find the expression for the backscattering current. We assume that the unperturbed density matrix of the helical edge electrons is given by
\begin{equation}
    \rho_0 = \frac{\exp\left[-\frac{1}{T}\int dy \Psi^\dagger(y) \left(i \sigma_z v \partial_y - \frac{\sigma_z V}{2}-\mu\right)\Psi(y)\right]}{\mathrm{Tr}_\mathrm{e}\exp\left[-\frac{1}{T}\int dy \Psi^\dagger(y) \left(i \sigma_z v \partial_y - \frac{\sigma_z V}{2}-\mu\right)\Psi(y)\right]}.
\end{equation}
Here $\mu$ is the chemical potential of the edge electrons, $V$ is the voltage applied to the helical edge, and $\mathrm{Tr}_\mathrm{e}$ is the trace over the states of the edge electrons. Note that while the density matrix $\rho_0$ is stationary, it describes a non-equilibrium situation with finite expectation of the edge spin density 
\begin{equation}
\langle s_j\rangle_0=\mathrm{Tr}_\mathrm{e}\left(\rho_0 \Psi^\dagger(y) (\sigma_j/2) \Psi(y) \right) = \delta_{jz}\nu V /2 .
\end{equation}
 To derive the quantum master equation for the reduced density matrix of the magnetic impurity we employ second order perturbation theory in the electron-impurity coupling constants $\mathcal{J}_{ij}$. To this end, we first decompose the electron-impurity interaction into a mean-field part and an ``irreducible'' part:
\begin{gather}\label{sm:e-i-h}
    \notag
    H_{\mathrm{e-i}}=\frac{\mathcal{J}_{ij}}{\nu}S_i s_j(y_0) = \frac{\mathcal{J}_{ij}}{\nu}S_i \langle s_j(y_0)\rangle_0 + \frac{\mathcal{J}_{ij}}{\nu}S_i \Bigl[s_j(y_0) \notag \\
    -\langle s_j(y_0)\rangle_0\Bigr] =\underbrace{\frac{V}{2}\mathcal{J}_{iz}S_i}_{H_{\mathrm{e-i}}^{\mathrm{mf}}} + \underbrace{\frac{\mathcal{J}_{ij}}{\nu} S_i :s_j(y_0):}_{H_{\mathrm{e-i}}^{\mathrm{irred}}}.
\end{gather}
Thus, the Hamiltonian of the whole system is given by
\begin{gather}
    H = \underbrace{iv\int dy \Psi^\dagger(y) \sigma_z\partial_y \Psi(y)}_{H_\mathrm{e}} + \underbrace{\mathcal{D}_{qp}S_qS_p + \frac{V}{2}\mathcal{J}_{iz} S_i}_{H_\mathrm{i}^{\mathrm{full}}}\notag \\
    +\underbrace{\frac{\mathcal{J}_{ij}}{\nu} S_i :s_j:}_{H_{\mathrm{e-i}}^{\mathrm{irred}}}.
    \label{sm:tot_h}
\end{gather}
We stress that $H_{\mathrm{i}}^{\mathrm{full}}=H_{\mathrm{i}}+H_{\mathrm{e-i}}^{\mathrm{mf}}$ contains no operators associated with the edge electrons. Next we introduce the joint density matrix of the impurity and the electrons: $\rho(t) = |\psi(t)\rangle\langle \psi(t)|$, where $|\psi(t)\rangle$ is the wave function of the whole system at  time $t$. The evolution of $\rho(t)$ is governed by the standard von-Neumann equation $d\rho(t)/dt = -i \left[H,\rho(t)\right]$. The goal of the subsequent derivation is to use this equation to extract the equation for the evolution of the reduced density matrix of the magnetic impurity, $\rho_S(t) = \mathrm{Tr}_\mathrm{e} \rho(t)$. First of all, we go to the interaction picture:
\begin{gather}
    \frac{d\rho_I(t)}{dt}=-i[\mathcal{V}_I(t),\rho_I(t)],\quad \rho(t)=U(t)\rho_I(t)U^{-1}(t),\notag \\
    U(t)=U_\mathrm{i}(t)U_{\mathrm{e}}(t)=U_{\mathrm{e}}(t)U_\mathrm{i}(t), \notag \\
     \mathcal{V}_I(t)=U^{-1}(t)H_{\mathrm{e-i}}^{\mathrm{irred}}U(t),\quad U_{\mathrm{e}}(t)=\exp\left(-iH_\mathrm{e}t\right),\notag \\
      U_\mathrm{i}(t)=\exp\left(-i H_{\mathrm{i}}^{\mathrm{full}}t\right).\label{sm:dm-int}
\end{gather}
In order to make the perturbative treatment possible we formally solve the evolution equation \eqref{sm:dm-int} and substitute the result back into \eqref{sm:dm-int}:
\begin{equation}
    \frac{d\rho_I}{dt}=-i[\mathcal{V}_I(t),\rho_I(-\infty)]+\int\limits_{-\infty}^{t} dt^\prime \left[\mathcal{V}_I(t),\left[\rho_I(t^\prime),\mathcal{V}_I(t^\prime)\right]\right].
\end{equation}
Tracing out electrons, we obtain
\begin{align}
    \frac{d\rho_{S,I}(t)}{dt}= & -i\mathrm{Tr}_\mathrm{e}\left[\mathcal{V}_I(t),\rho_I(-\infty)\right] 
    \notag \\
    & +\int\limits_{-\infty}^{t} dt^\prime \mathrm{Tr}_\mathrm{e}\Big(\left[\mathcal{V}_I(t),\left[\rho_I(t^\prime),\mathcal{V}_I(t^\prime)\right]\right]\Big),
\end{align}
where $\rho_{S,I}(t)= U_\mathrm{i}(t)\rho_S(t)U_\mathrm{i}^{-1}(t)$.
We assume that the electron-impurity interaction is switched on adiabatically, so that the distribution of the edge electrons is unperturbed at $t=-\infty$. Therefore, $\mathrm{Tr}_\mathrm{e}\left[\mathcal{V}_I(t),\rho_I(-\infty)\right]=0$, as $\mathcal{V}$ contains only irreducible electron operators. Moreover, in the weak coupling regime, $J\ll 1$, it is possible to approximately write $\rho_I(t) = \rho_{S,I} \otimes \rho_0$ on the right-hand side of the master equation \cite{Rivas}. Finally, substituting the explicit form of the perturbation $\mathcal{V}$ we find
\begin{gather}\notag
    \frac{d\rho_{S,I}}{dt}=\mathcal{J}_{rj}\mathcal{J}_{lk} \int\limits_{-\infty}^{t} dt^\prime \Big( \mathcal{K}_V^{jk}(t-t^\prime)\left[S_r^I(t^\prime)\rho_{S,I}(t^\prime),S_l^I(t)\right] \notag \\
    +\mathrm{h.c.}\Big), 
\end{gather}
where
\begin{gather}
\mathcal{K}_V^{jk}(\tau) = \frac{1}{\nu^2}\mathrm{Tr}_\mathrm{e}\left(\rho_0 :s_k^I(y_0,\tau) ::s_j^I(y_0,0):\right),\notag \\
 s_k^I(y_0,\tau)=U_\mathrm{e}^{-1}(\tau)s_k(y_0)U_\mathrm{e}(\tau),\notag \\
  S_l^I(t) = U_\mathrm{i}^{-1}(t)S_lU_\mathrm{i}(t).
\end{gather}

Next we employ the Markov approximation, i.e., we change $\rho_{S,I}(t^\prime)$ to $\rho_{S,I}(t)$. This approximation is justified because the correlators $\mathcal{K}_V^{jk}$ decay over a time proportional to either $1/V$ or $1/T$, while the relaxation time of $\rho_{S,I}$ has an additional large factor of $\mathcal J^{-2}$. Switching back to the Heisenberg picture, we get
\begin{align}
    \frac{d\rho_{S}(t)}{dt}=& -i\left[H_{\mathrm{i}}^{\mathrm{full}},\rho_S(t)\right] \notag \\
     +& \mathcal{J}_{rj}\mathcal{J}_{lk} \int\limits_{0}^{+\infty} d\tau\Big( \mathcal{K}_V^{jk}(\tau)\left[S_r^I(-\tau)\rho_{S}(t),S_l\right] +\mathrm{h.c.}\Big).
\end{align}
Now we introduce the eigenstates $|\psi_c\rangle$ of the full impurity Hamiltonian, $H_\mathrm{i}^\mathrm{full}|\psi_c\rangle=E_c|\psi_c\rangle$. Here the index $c$ takes one of $2S + 1$ values. Then it is possible to decompose the spin operators as \[S_r = \sum_{cd}\mathcal{S}^{cd}_r,\quad\mathcal{S}^{cd}_r = |\psi_c\rangle\langle\psi_c|S_r|\psi_d\rangle\langle\psi_d|.\]
Defining $\omega_{cd}=E_d - E_c$ and introducing $\mathcal{K}_V^{jk}(\omega) = \int_0^{+\infty}d\tau e^{i\omega\tau}\mathcal{K}_V^{jk}(\tau)$, we obtain
\begin{align}
    \frac{d\rho_{S}(t)}{dt}=& -i\left[H_{\mathrm{i}}^{\mathrm{full}},\rho_S(t)\right] \notag \\
    +& \mathcal{J}_{rj}\mathcal{J}_{lk} \Big( \sum_{cd}\mathcal{K}_V^{jk}(\omega_{cd})\left[\mathcal{S}_r^{cd}\rho_{S}(t),S_l\right] +\mathrm{h.c.}\Big).
    \label{eq: master eqn}
\end{align}
In order to write down the final form of the master equation, we calculate the correlators $\mathcal{K}_V^{jk}(\omega)$. This yields
\begin{align}
    \mathcal{K}_V^{jk}(\omega)= & \frac{i}{4}\sum_{\sigma_1,\sigma_2}\int d\xi_1d\xi_2\: \sigma_k^{\sigma_1\sigma_2} \sigma_j^{\sigma_2\sigma_1} \frac{1-n_F(\xi_2 -\frac{\sigma_2 V}{2})}{\omega+\xi_1-\xi_2+i0}
    \notag \\
    & \times n_F(\xi_1 - {\sigma_1 V}/{2}) ,
\end{align}
where $n_F(\varepsilon)= 1/[e^{(\varepsilon - \mu)/T}+1]$.

The correlator can be split into a Hermitian and an antihermitian parts:
\begin{equation}
    \mathcal{K}^{jk}_V(\omega)=\frac{1}{2}\mathcal{T}_V^{jk}(\omega)+i \mathcal{Q}^{jk}_V(\omega),\quad\mathcal{T}_V=\mathcal{T}_V^\dagger,\quad\mathcal{Q}_V=\mathcal{Q}_V^\dagger,
  \end{equation}
  where
  \begin{gather}  
    \mathcal{T}_V^{jk}(\omega)=\frac{\pi}{2}\sum_{\sigma_1,\sigma_2}\int d\xi_1d\xi_2\: \sigma_k^{\sigma_1\sigma_2} \sigma_j^{\sigma_2\sigma_1}\delta(\omega+\xi_1-\xi_2) 
    \notag \\
    \times (1-n_F(\xi_2 -\sigma_2 V/2))n_F(\xi_1 - \sigma_1 V/2),
    \end{gather}
    and
    \begin{gather}
    \mathcal{Q}_V^{jk}(\omega)=\frac{1}{4}\sum_{\sigma_1,\sigma_2}\mathrm{p.v.}\int d\xi_1d\xi_2\: \sigma_k^{\sigma_1\sigma_2} \sigma_j^{\sigma_2\sigma_1} 
    n_F \Bigl(\xi_1 - \frac{\sigma_1 V}{2} \Bigr)
    \notag \\
    \times \frac{(1-n_F(\xi_2 -\sigma_2 V/2))}{\omega+\xi_1-\xi_2},
\end{gather}
where $\mathrm{p.v.}$ denotes the Cauchy principal value.
$\mathcal{Q}_{jk}$ contains only logarithmically and linearly diverging (with the high energy cut-off $\sim |M|$) contributions.  The corresponding terms in the master equation \eqref{eq: master eqn} can be cast in the form of the unitary dynamics, i.e., they provide a renormalization of $H_{\rm i}^{\rm full}$.  
The logarithmically divergent contributions to $\mathcal{Q}_{jk}$ describe the Kondo renormalization (discussed in Appendix~\ref{app:sec:kondo}) of the coupling constants $\mathcal{J}_{jk}$ in $H_{\rm e-i}^{\rm mf}$. As we previously explained, we neglect the Kondo renormalization. The linearly diverging terms in $\mathcal{Q}_{jk}$ are consistent the with generation of the local anisotropy terms under the course of renormalization group flow in the Kondo problem with anisotropic exchange interaction \cite{RKonig,Schiller}.  In Eq.~\eqref{eq: master eqn} the corresponding terms can be viewed as correction to the local anisotropy Hamiltonian $H_{\rm i}$. However, the local anisotropy generated in this way due to edge states in parametrically smaller (it does not contain the large parameter $\Lambda_0^{\rm bulk}$) than the bulk contribution. Therefore, we can safely neglect it.

Tossing out $\mathcal{Q}_{jk}$, we finally obtain the quantum master equation 
in the form of Eq. \eqref{eq:denmat-ev}. The explicit calculation of the Hermitian part $\mathcal{T}_V^{jk}$ of the correlator matrix $\mathcal{K}^{jk}_V$ shows that $\mathcal{T}_V(\omega)=\mathcal{T}_V^+(\omega) + \mathcal{T}_V^-(\omega)$, where $\mathcal{T}_V^\pm(\omega)$ are given by Eq. \eqref{eq:Tm:def}.

The master equation allows us to find the reduced density matrix $\rho_S$ in the steady state. The next step is to employ this density matrix to evaluate the backscattering current mediated by the magnetic impurity. Once again, we switch to the interaction picture and, using Eq. \eqref{eq:dI:def}, find
\begin{align}
    \Delta I & = \mathrm{Tr} \left( i\left[H,\int dy s_z(y)\right] \rho(t)\right) \notag \\
    & = -\mathrm{Tr} \left( \frac{\mathcal{J}_{ir}}{\nu}S_i \varepsilon_{rzj} :s_j(y_0): \rho(t)\right) \notag \\
    & = -\mathrm{Tr} \left( \frac{\mathcal{J}_{ir}}{\nu}S_i^I(t) \varepsilon_{rzj} :s_j^I(y_0,t): \rho_I(t)\right).
\end{align}
Substituting the formal solution of the von Neumann equation into the expression above we obtain
\begin{gather}
    \Delta I  =i\frac{\mathcal{J}_{ir}\mathcal{J}_{lk}}{\nu^2}\varepsilon_{rzj}\int_{-\infty}^{t}dt^\prime \mathrm{tr}\Bigl (S_i^I(t):s_j^I(y_0,t):
    \notag \\
     \times \left[S_l^I(t^\prime):s_k^I(y_0,t^\prime):,\rho_I(t^\prime)\right]\Bigr).
\end{gather}
The subsequent calculations are similar to those in the derivation of the master equation. As a result, we find Eq. \eqref{eq:curr:corr-ex}.


\end{document}